\documentclass[twocolumn,twocolappendix]{aastex701}

\graphicspath{{./}{figures/}}
\makeatletter
\def\input@path{{tables/}{figures/}}
\makeatother

\usepackage{xspace}
\usepackage{mhchem}

\newenvironment{rotatepage}%
{\clearpage\pagebreak[4]\global\pdfpageattr\expandafter{\the\pdfpageattr/Rotate 90}}%
{\clearpage\pagebreak[4]\global\pdfpageattr\expandafter{\the\pdfpageattr/Rotate 0}}%

\newcommand{\Kepler}{\textit{Kepler}\xspace}

\newcommand{\Gaia}{\textit{Gaia}\xspace}

\newcommand{\Exofast}{\texttt{EXOFASTv2}\xspace}

\newcommand{\TTVFast}{\texttt{TTVFast}\xspace}

\newcommand{\bjdtdb}{\ensuremath{\mathrm{BJD}_\mathrm{TDB}}\xspace}
\newcommand{\Rstar}{\ensuremath{R_{\star}}\xspace} 
\newcommand{\Mstar}{\ensuremath{M_{\star}}\xspace}
\newcommand{\rhostar}{\ensuremath{\rho_{\star}}\xspace}
\newcommand{\Rjup}{\ensuremath{R_\mathrm{J}}\xspace} 
\newcommand{\Mjup}{\ensuremath{M_\mathrm{J}}\xspace}

\newcommand{\Rsun}{\ensuremath{R_\odot}\xspace} 
\newcommand{\Msun}{\ensuremath{M_\odot}\xspace}
\newcommand{\Rp}{\ensuremath{R_p}\xspace}
\newcommand{\Mp}{\ensuremath{M_p}\xspace}

\newcommand{\feh}{\ensuremath{\mathrm{[Fe/H]}}\xspace}

\newcommand{\jit}{\ensuremath{\sigma_{\mathrm{jit}}\xspace}}

\newcommand{\Porb}{\ensuremath{P_{\mathrm{orb}}\xspace}}
\newcommand{\ecosw}{\ensuremath{e\cos{\omega}\xspace}}
\newcommand{\esinw}{\ensuremath{e\sin{\omega}\xspace}}
\newcommand{\secosw}{\ensuremath{\sqrt{e}\cos{\omega}\xspace}}
\newcommand{\sesinw}{\ensuremath{\sqrt{e}\sin{\omega}\xspace}}
\newcommand{\imut}
{\ensuremath{i_\mathrm{mut}\xspace}}

\newcommand{\ms}{\ensuremath{\mathrm{m}\,\mathrm{s}^{-1}}\xspace}

\newcommand{\kms}{\ensuremath{\mathrm{km}\,\mathrm{s}^{-1}}\xspace}
\newcommand{\gcc}{\ensuremath{\mathrm{g}\,\mathrm{cm}^{-3}}\xspace}
\newcommand{\peryr}{\ensuremath{\mathrm{yr}^{-1}}\xspace}

\newcommand{\ticnum}{204698337}

\begin{document}

\title{A TESS Disappearing Act: The Vanishing Transits of TOI-2535\,b}

\author[0000-0001-7961-3907]{Samuel~W.~Yee}
\email[show]{syee@astro.ucla.edu}
\altaffiliation{51 Pegasi b Fellow}
\affiliation{Department of Physics \& Astronomy, University of California Los Angeles, Los Angeles, CA 90095, USA}

\author[0000-0003-0967-2893]{Erik~A.~Petigura}
\email{petigura@astro.ucla.edu}
\affiliation{Department of Physics \& Astronomy, University of California Los Angeles, Los Angeles, CA 90095, USA}

\author[0000-0002-9158-7315]{Rafael~Brahm}
\email{rafael.brahm@uai.cl}
\affiliation{Facultad de Ingeniería y Ciencias, Universidad Adolfo Ibáñez, Av. Diagonal las Torres 2640, Peñalolén, Santiago, Chile}

\author[0000-0001-6588-9574]{Karen~A.~Collins}
\affiliation{Center for Astrophysics \textbar Harvard \& Smithsonian, 60 Garden Street, Cambridge, MA 02138, USA}
\email{karen.collins@cfa.harvard.edu}

\author{Gregor Srdoc}									%
\affiliation{Kotizarovci Observatory, Sarsoni 90, 51216 Viskovo, Croatia}
\email{gregorsrdoc@gmail.com}

\author[0000-0001-9911-7388]{David W. Latham}
\affiliation{Center for Astrophysics \textbar Harvard \& Smithsonian, 60 Garden Street, Cambridge, MA 02138, USA}
\email{dlatham@cfa.harvard.edu}

\author[0000-0001-6637-5401]{Allyson~Bieryla}
\affiliation{Center for Astrophysics \textbar Harvard \& Smithsonian, 60 Garden Street, Cambridge, MA 02138, USA}
\email{abieryla@cfa.harvard.edu}

\author[0000-0003-0741-7661]{Victoria DiTomasso}
\affiliation{Center for Astrophysics \textbar Harvard \& Smithsonian, 60 Garden Street, Cambridge, MA 02138, USA}
\email{victoria.ditomasso@cfa.harvard.edu}

\author{Thomas Henning}
\affiliation{Max-Planck-Institut für Astronomie, Königstuhl 17, 69117 Heidelberg, Germany}
\email{henning@mpia.de }

\author[0000-0002-5389-3944]{Andr\'es Jord\'an} 
\affiliation{Facultad de Ingeniería y Ciencias, Universidad Adolfo Ibáñez, Av. Diagonal las Torres 2640, Peñalolén, Santiago, Chile}
\affil{Departamento de Astronom\'ia, Universidad de Chile, Casilla 36-D, Santiago, Chile}
\affil{El Sauce Observatory --- Obstech, Coquimbo, Chile}
\email{ajordan@astrofisica.cl}

\author[0000-0002-0236-775X]{Trifon Trifonov} 
\affiliation{Landessternwarte, Zentrum f\"ur Astronomie der Universit\"at Heidelberg, K\"onigstuhl 12, 69117 Heidelberg, Germany}
\affiliation{Department of Astronomy, Sofia University St Kliment Ohridski, 5 James Bourchier Boulevard, BG-1164 Sofia, Bulgaria}
\email{trifonov@phys.uni-sofia.bg}

\author[0000-0003-1464-9276]{Khalid~Barkaoui}
\affiliation{Instituto de Astrof\'isica de Canarias (IAC), E-38205 La Laguna, Tenerife, Spain}
\affiliation{Astrobiology Research Unit, University of Li\`ege, All\'ee du 6 ao\^ut, 19, 4000 Li\`ege (Sart-Tilman), Belgium}
\affiliation{Department of Earth, Atmospheric and Planetary Science, Massachusetts Institute of Technology, Cambridge, MA 02139, USA}
\email{Khalid.Barkaoui@uliege.be}
\author[0000-0002-5674-2404]{Phil Evans}
\affiliation{El Sauce Observatory, Coquimbo, 1870000, Chile}
\email{phil@astrofizz.com}
\author[[0000-0003-3986-0297]{Mourad~Ghachoui}
\affiliation{Cadi Ayyad University, Oukaimeden Observatory, High Energy Physics, Astrophysics and Geoscience Laboratory, Faculty of Sciences Semlalia, Marrakech, Morocco}
\affiliation{Astrobiology Research Unit, University of Li\`ege, All\'ee du 6 ao\^ut, 19, 4000 Li\`ege (Sart-Tilman), Belgium}
\email{mr.ghachoui@gmail.com}
\author[0000-0003-1462-7739]{Michaël~Gillon}
\affiliation{Astrobiology Research Unit, University of Li\`ege, All\'ee du 6 ao\^ut, 19, 4000 Li\`ege (Sart-Tilman), Belgium}
\email{michael.gillon@uliege.be}
\author{Eric Girardin}
\affiliation{Grand Pra Observatory, Switzerland}
\email{ergir@bluewin.ch}
\author[0000-0001-8923-488X]{Emmanuel Jehin}
\affiliation{Space Sciences, Technologies and Astrophysics Research (STAR) Institute, Université de Liège, Allée du 6 Août 19C, B-4000 Liège, Belgium}
\email{ejehin@uliege.be}
\author[0000-0001-9087-1245]{Felipe Murgas}
\affiliation{Instituto de Astrof\'isica de Canarias (IAC), E-38205 La Laguna, Tenerife, Spain}
\affiliation{Departamento de Astrof\'isica, Universidad de La Laguna (ULL), E-38206 La Laguna, Tenerife, Spain}
\email{fmurgas@iac.es}
\author[0000-0003-0987-1593]{Enric Palle}
\affiliation{Instituto de Astrof\'isica de Canarias (IAC), E-38205 La Laguna, Tenerife, Spain}
\affiliation{Departamento de Astrof\'isica, Universidad de La Laguna (ULL), E-38206 La Laguna, Tenerife, Spain}
\email{epalle@iac.es}
\author[0000-0003-3184-5228]{Adam~Popowicz}
\affiliation{Silesian University of Technology, Department of Electronics, Electrical Engineering and Microelectronics, Akademicka 16, 44-100 Gliwice, Poland}
\email{apopowicz@polsl.pl}
\author[0000-0002-3940-2360]{Don~J.~Radford}
\affiliation{Brierfield Observatory, New South Wales, Australia}
\email{dradford63@gmail.com}
\author{Howard M. Relles}
\affiliation{Center for Astrophysics \textbar Harvard \& Smithsonian, 60 Garden Street, Cambridge, MA 02138, USA}
\email{rellesh@yahoo.com}
\author[0000-0001-8227-1020]{Richard P. Schwarz}
\affiliation{Center for Astrophysics \textbar Harvard \& Smithsonian, 60 Garden Street, Cambridge, MA 02138, USA}
\email{rpschwarz@comcast.net}
\author[0000-0002-1836-3120]{Avi Shporer}
\affiliation{Department of Physics and Kavli Institute for Astrophysics and Space Research, Massachusetts Institute of Technology, Cambridge, MA 02139, USA}
\email{shporer@mit.edu}
\author[0000-0002-0345-2147]{Abderahmane Soubkiou}
\affiliation{Astrobiology Research Unit, University of Li\`ege, All\'ee du 6 ao\^ut, 19, 4000 Li\`ege (Sart-Tilman), Belgium}
\email{abdousoubkiou@gmail.com}
\author[0000-0002-3555-8464]{David~Watanabe}
\affiliation{Planetary Discoveries, Fredericksburg, VA 22405, USA}
\email{davidwatanabeastronomy@gmail.com}
\author[0000-0003-2127-8952]{Francis P. Wilkin} 
\affiliation{Department of Physics and Astronomy, Union College, 807 Union Street, Schenectady, NY 12308, USA}
\email{wilkinf@union.edu}
\author[0000-0002-0619-7639]{Carl~Ziegler}
\affiliation{Department of Physics, Engineering and Astronomy, Stephen F. Austin State University, 1936 North Street, Nacogdoches, TX 75962, USA}
\email{Carl.Ziegler@sfasu.edu}

\begin{abstract}
Planetary transits require a fortuitous alignment between a planet’s orbit and our line of sight, but dynamical interactions can render this geometric coincidence only temporary.
We report the discovery of the hot Jupiter TOI-2535\,b, the first confirmed planet orbiting a single star whose transits have been directly observed to disappear.
Over seven years of photometry from NASA's Transiting Exoplanet Survey Satellite (TESS), the planet's transit chord traversed and eventually moved off the stellar disk entirely, with an impact parameter changing at the rate of $\dot{b} = 0.16\,\peryr$.
The transit mid-times also show significant departures of $\pm 6$~hours from a linear ephemeris.
Joint modeling of the transit timing and duration variations (TTVs and TDVs) and NEID radial-velocity data indicates the presence of a nearby, mutually inclined ($\imut \gtrsim 5^\circ$) planetary perturber, with the TDV being primarily due to precession of the planets' orbital planes.
Our dynamical modeling reveals that the perturber is likely of Saturn-mass or greater and in a mean-motion resonance with TOI-2535\,b, making this an unusual system that contains two short-period giant planets.
The TDVs measured for TOI-2535\,b
represent one of the most extreme examples of planet-planet interactions driving orbital evolution on human timescales.
\end{abstract}

\keywords{\uat{Exoplanet astronomy}{486} --- \uat{Exoplanets}{498} --- \uat{Hot Jupiters}{753} --- \uat{Transit timing variation method}{1710} --- \uat{Transit duration variation method}{1707}}

\section{Introduction}

\begin{figure*}
\includegraphics[width=\linewidth]{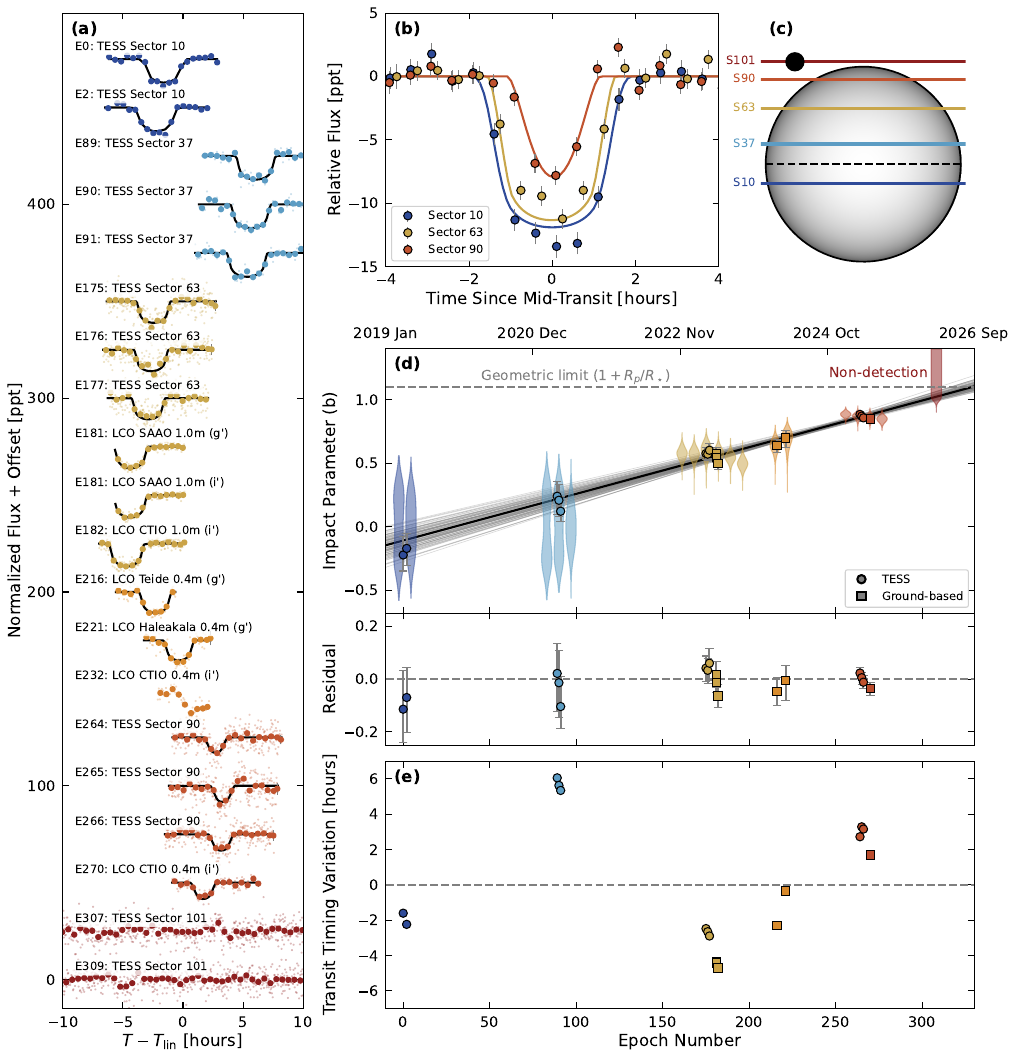}
\caption{\textbf{(a):} TESS and ground-based transit observations of TOI-2535\,b, phased to the best-fit linear ephemeris. Larger circles show the data binned to 1800s cadence. Transits are arranged chronologically with the earliest at the top, labeled by their epoch number. The best-fit model from our light curve fitting (\S\ref{sec:light_curve_analysis}) is shown in black. For reference, data from TESS Sector 101 are shown, with no obvious transit events throughout the full range of orbital phases spanned by the observed TTV amplitude.
\textbf{(b):} TESS transits from sectors 10, 63, and 90. Each sector's observations have been phase-folded and shifted to the best-fit local ephemeris, removing the TTVs, and binned to 1800s.
\textbf{(c):} Schematic showing the movement of the transit chord across the stellar disk. Each line represents the median impact parameter measured in each TESS sector. The dark circle shows the relative size of the planet to the star.
\textbf{(d):} Measured impact parameters for each transit. The violin plots show the posterior distribution of $b$ at each epoch, with the width representing the probability density. For clarity, the individual distributions have been offset horizontally.  The solid black line shows the median result from fitting the inferred impact parameters with a linear trend (\S\ref{ssec:nodal_precession}), while the light gray lines show 100 draws from the posterior distribution for this linear fit. The points with error bars show the median and 1$\sigma$ uncertainties on the \textit{unsigned} impact parameter $|b|$, under the assumption that the transit chord in Sector 10 is on the opposite hemisphere from those in the later sectors, as implied by the linear model.
\textbf{(e):} Extracted transit times of TOI-2535\,b, after subtracting the best-fit linear ephemeris. Note that the per-transit timing uncertainties (2--3~minutes) are too small to be visible on this scale.
\label{fig:combined_panel}}
\end{figure*}

In multi-planet systems, time-varying gravitational interactions between planets drive changes in the planets' orbital elements.
For planets that transit their host stars, these deviations from a Keplerian orbit can be detected through changes in the transit mid-times (transit timing variations; TTVs) or shapes (transit duration variations; TDVs). For a review of TTVs and TDVs, see \cite{Agol2025}. Because a planet is only observed to transit when its orbit normal is within a few degrees of the sky plane, even small changes in a planet's orbital inclination can result in measurable changes in its transit shape.
For sufficiently large inclination changes, a planet may only have the correct transit geometry for a fraction of its dynamical evolution; if the variation is fast enough and we are lucky to catch the planet in the right phase of its inclination evolution, we may see its transits appear or disappear.

Circumbinary planets frequently exhibit rapid changes in their transit impact parameters \citep[e.g.,][]{Kepler16_Doyle2011,Kepler35_Welsh2012,Kepler413_Kostov2014}, because the planet's orbit normal is precessing around the stellar binary's orbit normal, and because the stars themselves present a ``moving target'' for the planetary transit.
As a result, even a small misalignment between the two orbits leads to intervals when the planet may transit both, one, or neither of its host stars.

Around single stars, inclination variations of planets are typically less dramatic and appear to be relatively rare. For example, \citet{Shahaf2021} found only 15 Kepler Objects of Interest showing significant long-term TDVs from a search of 561 systems.
Still, the cases where TDVs are detected can provide interesting constraints on mutual inclinations in a multi-planet system \citep[e.g.,][]{Kepler108_Mills2017,Kepler448_Kepler693_Masuda2017,Nabbie2025}.
Some cases for which large TDVs have been detected include the planets TOI-216\,b \citep{TOI216_Dawson2021,TOI216_McKee2023} and K2-146\,c \citep{K2-146_Hamann2019,K2-146_Lam2019}, whose transits went from grazing to fully transiting over a 4.5 and 3 year observing baseline respectively.
Most recently, \citet{TOI4504_Almenara2026} reported the discovery of a new planet in the TOI-4504 system \citep{TOI4504_Vitkova2025}, which was non-transiting when the system was first observed, but entered a transiting configuration three years later.

These examples of planets whose orbits are precessing into view beg the complementary question---are there also planets whose transits are disappearing?\footnote{%
We note the case of KOI-120.01 \citep{KOI-120_Judkovsky2020}, a planet candidate from NASA's \Kepler mission \citep{Kepler_Borucki2010}, whose transits decreased in depth and vanished during the \Kepler observing baseline. However, the disposition of KOI-120.01 as a real planet is unclear given that the system contains at least two stars and the transit events may arise from an eclipsing binary.}
Some of the aforementioned systems showing large TDVs are expected to stop transiting on timescales of decades to centuries, but we have not yet caught a planet in the act of having its orbit tilt out of view.
NASA's ongoing Transiting Exoplanet Survey Satellite (TESS) mission \citep{TESS_Ricker15} provides an ideal dataset to search for such systems given its all-sky coverage and growing observational baseline, now approaching eight years for some systems. 

In this Letter, we present the discovery of TOI-2535\,b, a hot Jupiter with an orbital period of $\Porb = 8.236$~days that is the first planet directly observed to have ceased transiting.
Transits observed from TESS and with telescopes on the ground reveal that the planet's orbital inclination has changed so rapidly that the transit chord has likely traversed more than half the stellar disk, crossing the stellar equator before progressing to a grazing and finally non-transiting configuration, over the seven year observing baseline.
The transit mid-times also deviate significantly from a linear ephemeris, with TTVs of $\pm$6 hours, some of the largest ever seen for a short-period giant planet.
These transit variations indicate that TOI-2535\,b is in or near a mean-motion resonance with an inclined planetary companion, making it an unusual system  both in containing a hot Jupiter with a nearby planet, as well as in having two closely separated planets with significant mutual inclinations.

\section{Observations} \label{sec:observations}

\subsection{TESS Photometry} \label{ssec:photometry}

The TOI-2535 system (TIC \ticnum) was observed by TESS in five sectors between April 2019 and March 2026, first in the Full-Frame Images (FFIs) at 30-minute and 10-minute cadence in Sectors 10 and 37 respectively, and later in the ``postage stamps'' observed at 2-minute cadence in Sectors 63, 90, and 101.
A transiting planet candidate was first detected around TOI-2535 in a custom search of light curves by the Warm gIaNts with tEss (WINE) collaboration \citep{WINE_Brahm2019,wine2,wine3,wine4}, and announced as a community TESS Object of Interest (cTOI).
The candidate was later promoted to a TOI after vetting by the MIT TESS team \citep{TESS_cTOIs_Mireles2021}.\footnote{
In the TOI catalog, TOI-2535.01 is listed as having an orbital period of $\Porb = 16.4453$~days, twice that reported in this Letter. This is because one transit in the Sector 10 discovery data fell into a data gap, so the two transits observed in that sector are separated by $2\times\Porb$.}
We downloaded the light curves produced by the TESS SPOC pipeline \citep{TESS_SPOC_Jenkins2016,TESS_SPOC_Caldwell2020} for all sectors using the \texttt{lightkurve} package \citep{Lightkurve18}.\footnote{The TESS photometry of TOI-2535 can be obtained from the Mikulski Archive for Space Telescopes (MAST) via \dataset[doi:10.17909/5wpy-mg95]{\doi{10.17909/5wpy-mg95}}.}.

Inspection of the TESS data reveals that the transits are constant neither in period nor in shape over the seven years of TESS observations (Figure \ref{fig:combined_panel}).
The best-fit linear ephemeris has $\Porb = 8.236$~days, but individual transits occur early or late by several hours.
Furthermore, the transit event has decreased in depth and duration, such that by the final sector of TESS observations, no clear transit events are visible in the data.

\subsection{Ground-Based Follow-Up Observations} \label{ssec:ground_followup}

Following the identification of the transiting planet candidate orbiting TOI-2535, we began a follow-up campaign to observe the system with ground-based facilities to confirm its planetary nature, improve our characterization of the system, and to measure additional transit times and shapes to supplement the TESS observations.

\paragraph{Time-series Photometry}We obtained seeing-limited time-series photometry from several facilities with much higher angular resolution than TESS, with these observations being coordinated through the TESS Follow-up Observing Program Sub-Group 1 (TFOP SG1; \citealt{TFOP_Collins2018}).
The majority of the observations were obtained from the nodes of the Las Cumbres Observatory Global Telescope network \citep[LCOGT;][]{LCOGT_Brown2013} at Cerro Tololo Inter-American Observatory (CTIO), the South African Astronomical Observatory (SAAO), Teide Observatory, Haleakalā Observatory, and McDonald Observatory.
Additional observations came from the private El Sauce and Brierfield Observatories in Chile and Australia respectively, the TRAPPIST-South telescope at La Silla, Chile \citep{TRAPPIST_Jehin2011,TRAPPIST_Gillon2011}, and the Silesian University of Technology Observatories (SUTO) 0.3m telescope in Poland.
The data were reduced and light curves extracted using AstroImageJ \citep{AstroImageJ_Collins17}.
While we attempted a total of fifteen independent  observations from the ground, we successfully detected only six transits due to the uncertain ephemeris and large TTVs. 
Table \ref{tab:sg1_summary} in the appendices summarizes the ground-based photometric observations, while the successful transit detections are plotted in Figure \ref{fig:combined_panel}.

\paragraph{High Angular Resolution Imaging}We also observed TOI-2535 using the High-Resolution Camera, a speckle imaging instrument on the Southern Astrophysical Research (SOAR) 4.1m telescope (SOAR HRCam; \citealt{SOAR_Tokovinin2008}).
These observations were carried out and reduced using the standard observation strategy for TESS follow-up described in \citet{SOAR_TESS_Ziegler2019,SOAR_TESS_Ziegler2021}.
This allowed us to check for even closer stellar companions that could not be resolved by seeing-limited observations (Fig. \ref{fig:soar_imaging})

\paragraph{Reconnaissance Spectroscopy}
We first obtained a single reconnaissance spectrum of TOI-2535 with the Tillinghast Reflector Echelle Spectrograph (TRES; \citealt{TRES_Furesz2008}) on the 1.5m telescope at the Fred Lawrence Whipple Observatory (FLWO), with data reduced using the pipeline described in \citet{TRES_Buchhave2010}.
Two more spectra were observed with the Fiberfed Extended Range Optical Spectrograph (FEROS; \citealt{FEROS_Kaufer1999}) on the ESO 2.2m telescope at La Silla, Chile. FEROS data were processed with the automated \texttt{ceres} pipeline \citep{ceres}, which performs all the steps required to obtain RVs with the cross-correlation technique.

\paragraph{Precision Radial Velocities}
We subsequently observed TOI-2535 with the NEID spectrograph \citep{NEID_Schwab2016} on the WIYN 3.5m telescope at Kitt Peak National Observatory (KPNO), for the purpose of measuring precise radial velocities (RVs).
We used NEID in high-resolution (HR) mode and obtained a total of 15 spectra with 600s exposure times.
The data were reduced using v1.5.2 of the standard NEID Data Reduction Pipeline (NEID-DRP).\footnote{\url{https://neid.ipac.caltech.edu/docs/NEID-DRP}}
The NEID-DRP extracts precise RVs by cross-correlating each order with a weighted stellar line mask \citep{Baranne1996,Pepe2002}.
The per-order cross-correlation functions (CCFs) are then combined in a weighted sum and a final RV is extracted by fitting a Gaussian to the combined CCF.
The median RV precision achieved was 11.2~\ms, and the RV measurements are provided in Table \ref{tab:rv_obs}.

These ground-based follow-up observations confirmed that the observed transit events were on the target star; there are no signs of stellar binarity from the host star; and there is RV variation with $K \approx 100\,\ms$ at the linear ephemeris implied from the transit data, ruling out a stellar-mass companion. We therefore consider TOI-2535.01 confirmed as a \textit{bona fide} planet and henceforth refer to it as TOI-2535\,b, with further details and justification provided in Appendix \ref{app:planet_validation}.
We also used the ground-based data to characterize the host star, as described in Appendix \ref{app:stellar_characterization}. We report the derived stellar parameters in Table \ref{tab:stellar_params}, finding 
TOI-2535 to be an old, early K dwarf.
We also examined the 6D galactic kinematics of TOI-2535 and found that it is consistent with being part of the Milky Way's thick disk, which can be seen both by its position in the Toomre diagram (Figure \ref{fig:toomre}) and its moderately eccentric and inclined orbit within the galaxy.

\begin{deluxetable}{lcc}
\tablehead{
    \colhead{Parameter} & \colhead{Value} & \colhead{Source}
}
\tablecaption{TOI-2535 Stellar Parameters \label{tab:stellar_params}}
\startdata
TIC ID & 204698337 & TICv8 \\
Gaia DR3 ID & 3476765172730732032 & Gaia DR3 \\
2MASS ID & 12220715-2456349 & 2MASS \\
WISE ID & J122207.10-245635.3 & WISE \\
\multicolumn{3}{l}{\textbf{Gaia Astrometry}} \\
R.A. (J2016.0) & 12:22:07.07 & Gaia DR3 \\
Dec (J2016.0) & -24:56:35.49 & Gaia DR3 \\
$\mu_{\alpha*}$ (mas yr$^{-1}$) & $-69.074 \pm 0.016$ & Gaia DR3 \\
$\mu_{\delta}$ (mas yr$^{-1}$) & $-30.019 \pm 0.016$ & Gaia DR3 \\
$\varpi$ (mas) & $2.929 \pm 0.016$ & Gaia DR3 \\
RV (\kms) & $32.0 \pm 1.2$ & Gaia DR3 \\
RUWE & 1.12 & Gaia DR3 \\
\multicolumn{3}{l}{\textbf{Catalog Photometry}} \\
$T$ (mag) & $12.7024 \pm 0.0080$ & TICv8 \\
$G$ (mag) & $13.2475 \pm 0.0028$ & Gaia DR3 \\
$G_{BP}$ (mag) & $13.6976 \pm 0.0029$ & Gaia DR3 \\
$G_{RP}$ (mag) & $12.6350 \pm 0.0038$ & Gaia DR3 \\
$J$ (mag) & $11.902 \pm 0.024$ & 2MASS \\
$H$ (mag) & $11.453 \pm 0.021$ & 2MASS \\
$K_s$ (mag) & $11.381 \pm 0.023$ & 2MASS \\
W1 (mag) & $11.335 \pm 0.023$ & WISE \\
W2 (mag) & $11.386 \pm 0.022$ & WISE \\
W3 (mag) & $11.31 \pm 0.17$ & WISE \\
\multicolumn{3}{l}{\textbf{Spectroscopic Parameters from \texttt{SpecMatch-Emp}}} \\
$T_{\rm eff}$ (K) & $5180 \pm 110$ & this work \\
$\log{g}$ (cgs) & $4.50 \pm 0.12$ & this work \\
$\feh$ (dex) & $0.095 \pm 0.090$ & this work \\
\multicolumn{3}{l}{\textbf{Physical Stellar Parameters from \Exofast}} \\
$M_\star$ (\Msun) & $0.864^{+0.036}_{-0.028}$ & this work \\
$R_\star$ (\Rsun) & $0.877^{+0.028}_{-0.027}$ & this work \\
$\rho_\star$ (\gcc) & $1.81^{+0.19}_{-0.16}$ & this work \\
$T_{\rm eff}$ (K) & $5256 \pm 80$ & this work \\
$\log{g}$ (cgs) & $4.488^{+0.031}_{-0.029}$ & this work \\
$\feh$ (dex) & $0.101^{+0.053}_{-0.052}$ & this work \\
Age (Gyr) & $9.7^{+2.8}_{-4.0}$ & this work \\
$A_V$ (mag) & $0.242^{+0.018}_{-0.033}$ & this work%

\enddata
\tablerefs{Catalog astrometry and photometry are drawn from \Gaia DR3 \citep{GaiaDR3_Vallenari2022}, 2MASS \citep{TMASS_Cutri2003,TMASS_Skrutskie2006}, WISE \citep{WISE_Cutri2012}, and the TESS Input Catalog \citep{TIC_Stassun2018,TIC_Stassun2019}.}
\end{deluxetable}

\section{Light Curve Analysis}
\subsection{Transit Timing and Shape Modeling \label{sec:light_curve_analysis}}

In this section, we measure the timing and shape variations of TOI-2535\,b's transits.
Although TDVs can arise from changes in the planet's orbital velocity due to changes in eccentricity or apsidal precession \citep[e.g. KOI-142;][]{KOI-142_Nesvorny2013}, the TDVs driven by these phenomena are expected to be $\mathcal{O}(a/\Rstar)$ smaller than those caused by inclination variations \citep{Dawson2020,Shahaf2021}. We thus assume that the observed TDVs are due to changes in orbital inclination, or equivalently, impact parameter $b$.

We used the differentiable transit model of \citet{exoplanet_Agol2020} as implemented in the \texttt{exoplanet} code \citep{exoplanet_ForemanMackey2021} to model the photometry. We used the No U-Turn Sampler (NUTS; \citealt{NUTS_Hoffman})  to efficiently explore the parameter posterior distributions.
We fitted the data treating each transit observation as a separate light curve. 
For the TESS observations, we extracted the light curve within $\pm$ 3 transit durations of each transit event and treated them as separate observations.
We excluded the LCO 2024 Jun 25 observation (Epoch 232) from our final fit as no transit egress was observed in that sequence, limiting the precision to which the transit time and duration could be measured. We also did not include the TESS Sector 101 data in this fit (see \S\ref{ssec:nondetection}).

The parameterization and priors used are detailed in Appendix \ref{app:transit_fit_params}, but we summarize the key points here.
We parameterized the transit model using the full seven-parameter basis (\Porb, $T_c$, $r\equiv \Rp/\Rstar$, \rhostar, $b$, $\secosw$, $\sesinw$), avoiding implicit biases that may arise from alternate parameterizations that use the stellar pseudodensity \citep{Dawson2020,Gilbert2022}.
The following parameters were shared across all transit observations: \Porb, $r$, $\rhostar$, $e$, and $\omega$; while we fitted separate $b$ and $T_c$ for each transit observation to account for the TTVs and TDVs.
While transit observations only constrain the unsigned impact parameter $|b|$, we fitted for the signed value of $b$ for the TESS Sector 10 and 37 data, to avoid sampling issues near the boundary at $b = 0$. This also allowed us to be agnostic to which hemisphere of the star the transit chord is on (see \S\ref{ssec:nodal_precession}).
Each transit was fit simultaneously with a detrending model to account for potential systematics. For the ground-based data, the detrending model was a linear function of the vectors specified in Table \ref{tab:sg1_summary}, while for the TESS data, we used a quadratic function of time.

The detrended data and best-fitting transit models are plotted in panel (a) of Figure \ref{fig:combined_panel}.
The transit times and measured impact parameters for each transit, along with 1-$\sigma$ confidence intervals, are reported in Table \ref{tab:transit_times} and plotted in panels (d) and (e) of Figure \ref{fig:combined_panel}.
The inferred impact parameters of TOI-2535\,b steadily increased over the observation baseline, from a nearly equatorial transit chord to one that is highly grazing.
The transit times show excursions from a linear ephemeris with a maximum amplitude of 6 hours.
We performed a preliminary fit of the TTVs using a sinusoidal model, as a single frequency typically dominates the TTV signal \citep[e.g.,][]{Lithwick2012}; however, we found that no single sinusoid could fit all the data.
A two-sinusoid model provided a more reasonable fit to the TTV measurements, but with our transit timing measurements clustered around eight well-separated epochs, many different period aliases are plausible.

\begin{deluxetable}{cccc}
\tablecolumns{4}
\tablecaption{Measured Mid-Transit Times and Impact Parameters \label{tab:transit_times}}
\tablehead{
    \colhead{Epoch} & \colhead{Transit Time (\bjdtdb)\tablenotemark{a}} & \colhead{Impact Parameter $|b|$} & \colhead{Facility}
}
\startdata
0 & $2458576.7685(19)(20)$ & $0.22^{+0.13}_{-0.14}$ & TESS Sector 10 \\
2 & $2458593.2139(27)(23)$ & $0.17^{+0.07}_{-0.16}$ & TESS Sector 10 \\
89 & $2459310.0744(16)(18)$ & $0.24^{+0.13}_{-0.12}$ & TESS Sector 37 \\
90 & $2459318.2924(21)(17)$ & $0.21^{+0.12}_{-0.13}$ & TESS Sector 37 \\
91 & $2459326.5163(21)$ & $0.12^{+0.05}_{-0.12}$ & TESS Sector 37 \\
175 & $2460017.9982(17)(20)$ & $0.575^{+0.045}_{-0.049}$ & TESS Sector 63 \\
176 & $2460026.2266(18)(22)$ & $0.571^{+0.042}_{-0.053}$ & TESS Sector 63 \\
177 & $2460034.4524(25)(29)$ & $0.601^{+0.067}_{-0.054}$ & TESS Sector 63 \\
181 & $2460067.3348(11)(12)$ & $0.574^{+0.049}_{-0.046}$ & LCO SAAO 1.0m ($g^\prime$) \\
181 & $2460067.3323(11)(12)$ & $0.544^{+0.042}_{-0.037}$ & LCO SAAO 1.0m ($i^\prime$) \\
182 & $2460075.55599(89)(91)$ & $0.495 \pm 0.044$ & LCO CTIO 1.0m ($i^\prime$) \\
216 & $2460355.6733(19)$ & $0.637^{+0.059}_{-0.043}$ & LCO Teide 0.4m ($g^\prime$) \\
221 & $2460396.9339(36)(32)$ & $0.698^{+0.072}_{-0.055}$ & LCO Haleakala 0.4m ($g^\prime$) \\
264 & $2460751.2015(28)(29)$ & $0.884^{+0.022}_{-0.018}$ & TESS Sector 90 \\
265 & $2460759.4607(31)(27)$ & $0.870^{+0.025}_{-0.020}$ & TESS Sector 90 \\
266 & $2460767.6912(25)(30)$ & $0.856^{+0.027}_{-0.022}$ & TESS Sector 90 \\
270 & $2460800.5734(26)(28)$ & $0.847^{+0.023}_{-0.027}$ & LCO CTIO 0.4m ($i^\prime$)%
\enddata
\tablenotetext{a}{The numbers in parentheses represent the uncertainty in the last two decimal digits. If two sets of numbers are provided, these represent the upper and lower 1-$\sigma$ uncertainties respectively.}
\end{deluxetable}

\subsection{Sector 101 Non-Detection} \label{ssec:nondetection}

\begin{figure}
    \centering
    \includegraphics[width=\linewidth]{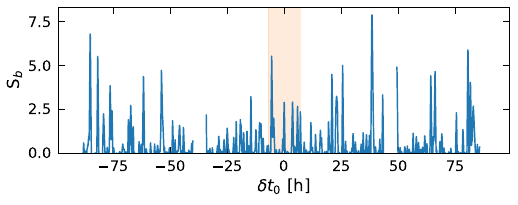}
    \includegraphics[width=\linewidth]{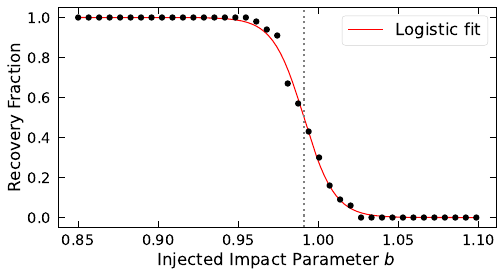}
    \caption{\textbf{Top:} Transit detection statistic $S_b$ as a function of phase offset relative to the time predicted by a linear ephemeris. We expect that any transits of TOI-2535\,b should have $|\delta T_c| < 6$~hours, corresponding to the shaded region. 
    \textbf{Bottom:} Recovery fraction of injected transits with different impact parameters (black points), along with a fit of a logistic curve to these points (red line).
    }
    \label{fig:s101_nondetection}
\end{figure}

In addition to the transit detections, we sought to quantify the non-detection of any transits in TESS Sector 101.
First, we performed a specialized search for transits in the light curve.
The planetary transit parameters are well-constrained from the other light curve data, including the orbital period, which varies by only a small fractional amount; thus, a conventional blind search would provide only a conservative limit on the transit detectability.
On the other hand, because of the large TTVs and uncertainty in the shape of the TTV curve, the time of conjunction at the epoch of the Sector 101 transits is unknown and remains a parameter to be searched over.

We constructed a set of transit templates using the best-fit planet properties from the previous section and listed in Table \ref{tab:transit_params}, spanning a range of impact parameters $b \in [0.85, 1+\Rp/\Rstar]$.
We define the signal detection statistic
\begin{equation}
S(b, \delta T_c) \equiv \chi^2_\mathrm{flat} - \chi^2_\mathrm{T(b)}(\delta T_c),
\end{equation}
the $\Delta \chi^2$ in fitting the Sector 101 light curve with the transit template $T(b)$ compared with a ``no-transit'' flat model, for different shifts in the time of conjunction $\delta T_c$ from the time predicted by the linear ephemeris.
We then take $S_b(\delta T_c)$ as the maximum statistic over all templates for each $\delta T_c$, obtaining an analogy to the standard periodogram, only as a function of the phase offset instead of orbital period (Figure \ref{fig:s101_nondetection}).

Given that the observed TTVs in the detected transits span a range of $\pm 6$~hours, we restrict our search to a conservative window in $\delta T_c$ of $\pm 12$~hours. We see in Figure \ref{fig:s101_nondetection} that the largest $S_b$ within this range does not exceed peaks outside the window, which are likely spurious and due to matches between the transit template and time-correlated noise. This is a strong indication that TOI-2535\,b did not transit in Sector 101, although the detection threshold needs to be calibrated to assess the statistic significance of this result.

We calibrate the detection statistic as follows: we take 
\begin{equation}
S_{\max} \equiv \max_{\delta T_c} S_b(\delta  T_c), 
\end{equation}
where the maximum is taken over the $-12\,\mathrm{hr} <\delta T_c < 12\,\mathrm{hr}$ window where we expect the planet's transits to be.
In the real data, this value is $S_{\max} = 5.18$.
We then construct synthetic light curves by selecting random 12 hour blocks from the real light curve outside the potential transit window, such that any time-correlated noise on shorter time scales is preserved. We then perform the same search for transits in these synthetic light curves and compute $S_{\max}$, repeating this 20{,}000 times.
This provides the null distribution of $S_{\max}$ when no transits are present, and we take the 99.7th percentile (3$\sigma$) as our detection threshold.
We find $S_{\max,\,\mathrm{threshold}} = 18.38$, confirming the non-detection.

We then investigate the minimum impact parameter TOI-2535\,b must have had to avoid detection in the Sector 101 data.
We injected transits into the light curve with varying impact parameters $b$ and $\delta T_c$ within the expected transit window.
A transit search as described above was carried out and we consider a transit recovered if the trial $S_{\max} > S_{\max,\,\mathrm{threshold}}$. 
The bottom panel of Figure \ref{fig:s101_nondetection} shows the recovery fraction as a function of $b$; we find that even transit events as grazing as $b = 0.991$ would have been recovered with 50\% probability.
We fitted a (flipped) logistic function to these recovery fractions; this fit is used to plot the ``non-detection'' probability density in panel (d) of Figure \ref{fig:combined_panel}.

\section{TTV and TDV Modeling}
\subsection{Nodal Precession Model} \label{ssec:nodal_precession}

Given the steady increase in impact parameters seen in Figure \ref{fig:combined_panel}, we decided to fit this variation with the simplest phenomenological model, an impact parameter changing linearly with time:
\begin{equation}
b(t) = \dot{b}\times(t-T_{c,0}) + b_0,
\end{equation}
where $T_{c,0}$ is the mid-transit time of transit $i = 0$.
Let $D_i$ denote the photometry associated with transit $i$, and $D=\{D_i\}$ is the full data set. Under standard Bayesian inference assumptions, the posterior probability distribution for the parameters $\{\dot{b}, b_0\}$ is proportional to the product of the prior $\pi$ and likelihood $\mathcal{L}_\mathrm{TDV}$:
\begin{equation}
p(\dot{b}, b_0\,|\,D ) \propto \pi(\dot{b}, b_0) \mathcal{L}_\mathrm{TDV}(D\,|\,\dot{b}, b_0). \label{eq:linear_posterior}
\end{equation}
Rather than assuming a standard Gaussian likelihood, we use the marginalized posterior density for each transit's $b_i$ from the light-curve fitting: 
\begin{equation}
\mathcal{L}_\mathrm{TDV}(D_i\,|\,b_i=\dot{b}t_i + b_0) \propto \frac{p(b_i\,|\,D_i)}{\pi(b_i)},
\end{equation}
where we approximated $p(b_i\,|\,D_i)$ from the posterior samples using a Gaussian Kernel Density Estimate (KDE).
In the light-curve fitting, we imposed uniform priors on $b_i$ so $\pi(b_i) \propto \mathrm{constant}$.
Assuming that the photometric data sets for the individual transits are conditionally independent, we can write the total likelihood as the product of the individual likelihoods for each transit,
\begin{equation} \label{eq:tdv_likelihood}
\mathcal{L}_\mathrm{TDV}(D\,|\,\dot{b}, b_0) = \prod_i \mathcal{L}(D_i\,|\,b_i) \propto \prod_i p(b_i\,|\,D_i).
\end{equation}
Inspection of the joint light curve posterior revealed no strong covariances between the inferred $b_i$, so this decomposition of the likelihood is reasonable.
We also incorporated the Sector 101 transit non-detection by assuming that $P(b_\mathrm{S101}\,|\,D_\mathrm{S101})$ is proportional to the logistic functional form fitted to the recovery fraction shown in Figure \ref{fig:s101_nondetection}.

Because we use a posterior density for the signed value of $b$ in the early sectors, this analysis is agnostic as to which hemisphere of the star the transit chord traverses.\footnote{
Because we enforced $b > 0$ for the data after TESS Sector 37, this requires the solution to have $\dot{b} > 0$; we cannot distinguish the symmetric solution with $(\dot{b}, b_0) \rightarrow (-\dot{b}, -b_0)$.}
We used the \texttt{emcee} Markov Chain Monte Carlo (MCMC) package \citep{Emcee_Foreman-Mackey13} to explore the posterior distribution for $(\dot{b}, b_0) $ given uniform priors in Equation \ref{eq:linear_posterior}.

The best-fitting linear model is shown in panel (d) of Figure \ref{fig:combined_panel}.
We find that the data are surprisingly well-described by a linear trend with $\dot{b} = 0.16\pm0.01\,\peryr$.
$97.6\%$ of posterior samples have $b_0 < 0$, i.e. over the observation baseline, the increase in impact parameter has taken the planet's transit chord from one hemisphere to the other, and eventually to a configuration where it no longer transits the face of its star.
This is the fastest rate of inclination change observed for an exoplanet orbiting a single star.

\begin{figure}
\centering
\includegraphics[width=\linewidth]{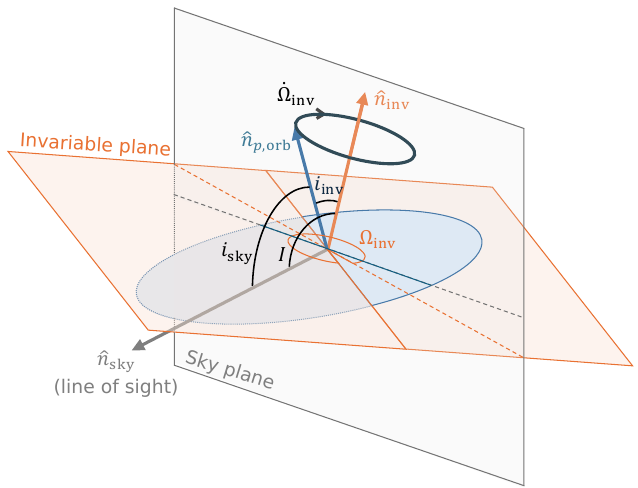}
\includegraphics[width=\linewidth]{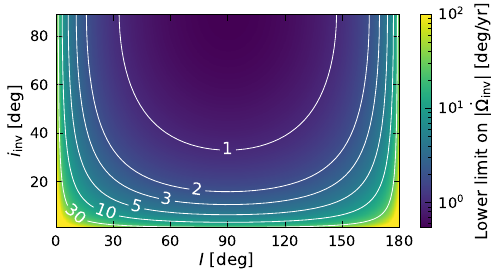}
\caption{\textbf{Top:} Schematic showing definition of angles used. 
$\Omega_\mathrm{inv}$ is the longitude of ascending node measured in the invariable plane, with the reference direction chosen to be along the intersection between the sky and invariable planes.
As the orbit normal precesses around the normal to the invariable plane, the planet's inclination with respect to the invariable plane $i_\mathrm{inv}$ remains fixed, but the observed sky-plane inclination $i_\mathrm{sky}$ varies.
\textbf{Bottom:}
Lower limit on $\dot{\Omega}_\mathrm{inv}$ from our measured $\dot{b}$, for a range of planetary inclinations $i_\mathrm{inv}$ and angles between the invariant and sky planes $I$. White lines show contours of constant $|\dot{\Omega}_\mathrm{inv}|$.
The configuration shown in the top panel corresponds to a transiting planet system with low mutual inclinations, such that $i_\mathrm{sky} \sim I \sim 90^\circ$, $i_\mathrm{inv} \lesssim 10^\circ$. In such a system, the typical lower limit on $|\dot{\Omega}_\mathrm{inv}|$ based on our measured $\dot{b}$ is an order of magnitude larger than the conservative limit of $> 0.5^\circ\,\peryr$.}
\label{fig:omega_dot_lower_limit}
\end{figure}

To interpret this rate of change in terms of the planet's orbital elements, we use the coordinate system shown in Figure \ref{fig:omega_dot_lower_limit}. We define $i_\mathrm{inv}$, $\Omega_\mathrm{inv}$ as the inclination and longitude of ascending node of TOI-2535\,b, measured with respect to the invariable plane. Similarly, $i_\mathrm{sky}$, $\Omega_\mathrm{sky}$ are the same quantities but measured relative to the sky plane.
Upper-case $I$ is the inclination between the invariable and sky planes.
Then the planet's inclination relative to the sky plane is given by
\begin{equation}
\cos{i_\mathrm{sky} (t)} = \cos{I}\cos{i_\mathrm{inv}} + \sin{I}\sin{i_\mathrm{inv}}\cos{\Omega_\mathrm{inv}}
\end{equation}
and its impact parameter is \citep{winn2010a}
\begin{equation}
b(t) = \frac{a}{R_\star} \frac{1-e^2}{1+e\sin{\omega}} \cos{i_{\mathrm{sky}}(t)}. 
\end{equation}

Defining $B \equiv \frac{a}{R_\star} \frac{1-e^2}{1+e\sin{\omega}}$, the rate of change in impact parameter is
\begin{equation}
\begin{split}
\dot{b}(t) =& -B\left[\frac{di_{\mathrm{inv}}}{dt}(\cos{I}\sin{i_{\mathrm{inv}}} - \sin{I}\cos{i_{\mathrm{inv}}}\cos{\Omega_{\mathrm{inv}}}) \right. \\
&\qquad\,\,\left. + \dot{\Omega}_{\mathrm{inv}}\sin{I} \sin{i_{\mathrm{inv}}} \sin{\Omega_{\mathrm{inv}}} \right] \\
& + \dot{B}\cos{i_\mathrm{sky}(t)}.
\end{split}
\label{eq:bdot}
\end{equation}
The first term in the square parentheses represents changes in the planet's inclination relative to the invariable plane, while the second is due to precession of the longitude of nodes at fixed inclination.
The final term is due to eccentricity variations and apsidal precession.
In principle, all three terms can cause significant changes to the impact parameter.

We first consider whether apsidal precession of an eccentric orbit alone, whether from tidal or multi-body interactions, could be responsible for the observed TDVs. Under the assumption of constant eccentricity and $i_\mathrm{sky}$, apsidal precession would cause a maximum variation of $b_{\max}/b_{\min} = (1+e)/(1-e)$.
Our measurements of $|b|$ from individual transits range from 0.2 to 1.1, which would require $e > 0.692$, a high eccentricity for such a short-period planet.
Furthermore, this would cause a transit whose duration is shorter than that expected for a planet with a circular orbit by a factor of $\sqrt{1-e^2}/(1 + e) \approx 0.43$ at $b = b_\mathrm{min}$.
This is ruled out by the actual measured transit duration, which is consistent with a circular orbit given the planet's orbital period and our knowledge of the stellar density (Table \ref{tab:transit_params}), as well as by the RV data.
For lower eccentricities, apsidal precession will cause only smaller changes in the impact parameter, so we turn to the other terms in Equation \ref{eq:bdot}.

Interactions with other bodies in the system can cause changes in both $i_\mathrm{inv}$ and $\Omega_\mathrm{inv}$. However,
we find in Section \ref{ssec:dynamical_analysis} that for plausible two-planet configurations that can explain both the TTVs and TDVs, the dominant contribution is the nodal precession term, whereas $i_\mathrm{inv}$ varies by only a small amount (see also Appendix \ref{app:ttv_model_details}).
On timescales short relative to the nodal precession timescale, $\sin{\Omega_{\mathrm{inv}}} \sim \mathrm{constant}$, so $\dot{b}$ is directly proportional to the precession rate $\dot{\Omega}_{\mathrm{inv}}$, so our measurement places a robust lower limit on this rate, $|\dot{\Omega}_{\mathrm{inv}}| > 0.5^\circ\,\peryr$, corresponding to a precession period $P_\Omega = 2\pi/\dot{\Omega}_{\mathrm{inv}} \lesssim 700$~yr.

Such rapid precession also allows us to exclude certain mechanisms that might cause nodal precession, such as interactions with the stellar quadrupole moment, as has been seen for several hot Jupiters orbiting early type stars \citep[e.g.,][]{Kepler13_Szabo2012,WASP33_Johnson2015,KELT9_Stephan2022}.
\citet{Miralda-Escude2002} derived the precession rate due to this effect; we find that a quadrupole moment of $J_2 \gtrsim 10^{-2}$ would  be required to cause $\dot{\Omega}_{\mathrm{inv}} > 0.5^\circ\,\peryr$; far larger than expected for an old, slowly rotating star.
This scenario is also disfavored from an Occam's razor perspective since the stellar quadrupole cannot drive sinusoidal TTVs.
As such, we exclude interactions with the stellar spin from subsequent analysis.

We also consider the possibility that the TTVs and TDVs are due to a hierarchical ($P_2 \gg P_1)$, highly eccentric perturber, similar to the architectures of the Kepler-419 \citep{Kepler419_Dawson2014c}, Kepler-448, and Kepler-693 systems \citep{Kepler448_Kepler693_Masuda2017}.
\citet{Borkovits2015} derived the nodal precession rate for such hierarchical systems (their equation C3)\footnote{Note that our $\Omega_{\mathrm{inv}}$ is equivalent to the ``dynamical node'' $h$ defined in \citet{Borkovits2015} and differs from their definition of $\Omega_1$.}:
\begin{equation}
    \frac{P_\mathrm{1}}{2\pi} \dot{\Omega}_{\mathrm{inv}} = -\frac{2}{5}A_\mathrm{secl} \frac{C}{C_2} \frac{\cos{i_\mathrm{mut}}}{(1-e_1^2)^{1/2}} \left[\left(1+\frac{3}{2}e_1^2\right) - \frac{5}{2}e_1^2\cos{2g_1}\right],
    \label{eq:borkovits_prec}
\end{equation}
where the subscripts (1, 2) refer to the inner and outer body, $C_2$ is the orbital angular momentum of the outer object while $C$ is the total orbital angular momentum, $g_1$ is the argument of periastron measured from the line of nodes in the invariable plane, and $A_\mathrm{sec1}$ is a dimensionless amplitude:
\begin{align}
\begin{split}
A_\mathrm{sec1} &= A_\mathrm{L1}\frac{P_1}{P_2} \\
&= \frac{15}{8} \frac{M_2}{M_1 +M_2+M_\star} \left(\frac{P_1}{P_2}\right)^2 (1 - e_2^2)^{-3/2}
\end{split}
\end{align}

If we assume that the eccentricity of the inner hot Jupiter is small and neglect the $e_1^2$ terms, we can write Equation \ref{eq:borkovits_prec} as:
\begin{equation}
\dot{\Omega}_{\mathrm{inv}}\sin{i_{\mathrm{inv}}} = -\frac{3\pi}{4} \frac{M_2}{M_\mathrm{tot}} \frac{P_1}{P_2^2} \frac{\sin{2i_\mathrm{mut}}}{\left(1-e_2^2\right)^{3/2}},
\end{equation}
where we have made use of the fact that $\sin{i_{\mathrm{inv}}} = (C_2/C)\sin{i_\mathrm{mut}}$. Using our measured lower limit of $|\dot{\Omega}_{\mathrm{inv}}\sin{i_{\mathrm{inv}}}| > 0.5^\circ\,\peryr$, we get the following constraint:
\begin{equation}
\frac{M_2}{M_\mathrm{tot}} >  0.16 \left(\frac{P_2}{\mathrm{yr}} \right)^2 \frac{\left(1-e_2^2\right)^{3/2}}{\sin{2i_\mathrm{mut}}}.
\label{eq:min_mass}
\end{equation}
This expression reveals that a hierarchical perturber would need to be either extremely massive or on a highly eccentric orbit to cause precession as rapid as observed.
Even in the limit of orbit-crossing eccentricity, the perturber would need to have $M_2 > 8.7\,\Mjup$ with $P_2 = 1$~yr, with this minimum mass growing approximately linearly with period.
Such massive perturbers are excluded by the lack of $\sim$km/s scale RV variations.
Furthermore, Equation (\ref{eq:min_mass}) shows that the projected precession rate is maximized when $\imut = 45^\circ$.
Thus, while a massive perturber on a close to face-on orbit (i.e., nearly perpendicular to that of TOI-2535\,b) could cause rapid precession while having a small RV signal, the observable precession rate, which is proportional to $\dot{\Omega}_\mathrm{inv}\sin{i}_\mathrm{inv}$, would be suppressed in such a configuration.

This leaves the most likely explanation for the TTVs and TDVs as a non-transiting neighboring planet in or close to a mean-motion resonance with TOI-2535\,b.
Depending on the proximity to resonance, this perturber could cause TTVs of multiple hours and rapid nodal precession.
In such a compact multi-planet system, one might expect low-to-moderate relative inclination between the planets' orbits.
For example, \citet{He2020} found that the distribution of multi-planet systems discovered by \textit{Kepler} could be modeled by a planet population with multiplicity-dependent but low intrinsic mutual inclinations, finding a median $i_\mathrm{mut} \sim 5^\circ$ for intrinsically two-planet systems.
If the planets in the TOI-2535 system have similarly low mutual inclinations, then this also requires that the invariant plane is close to perpendicular to the sky plane ($I \sim 90^\circ$).
Such a configuration would significantly increase the lower limit on $\dot{\Omega}_\mathrm{inv}$ by more than order of magnitude (Figure \ref{fig:omega_dot_lower_limit}). 
Indeed, we find in the next section that system architectures able to reproduce the observations have precession periods on the order of decades, rather than centuries.

\subsection{$N$-body Modeling} \label{ssec:dynamical_analysis}

\begin{figure*}
\centering
\includegraphics[width=\linewidth]{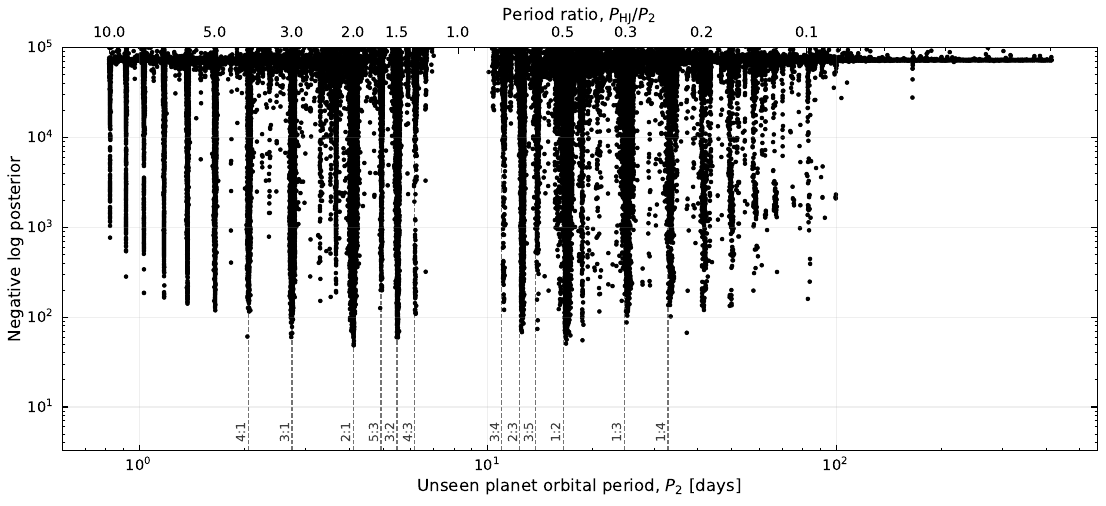}
\caption{Result of grid search and local optimization procedure used for finding plausible system architectures. We explored period ratios between the hot Jupiter and unseen perturber ranging from $P_\mathrm{HJ}/P_2 = 0.02-10$ and optimized a \texttt{TTVFast} model to fit the observed TTVs, TDVs, and RV data. No unique solution for the unseen perturber could be found, with the most likely families of solutions from our search clustered around the 2:1 and 3:2 period commensurabilities with the hot Jupiter.
\label{fig:ttv_search}}
\end{figure*}

\begin{figure*}
\centering
\includegraphics[width=\linewidth]{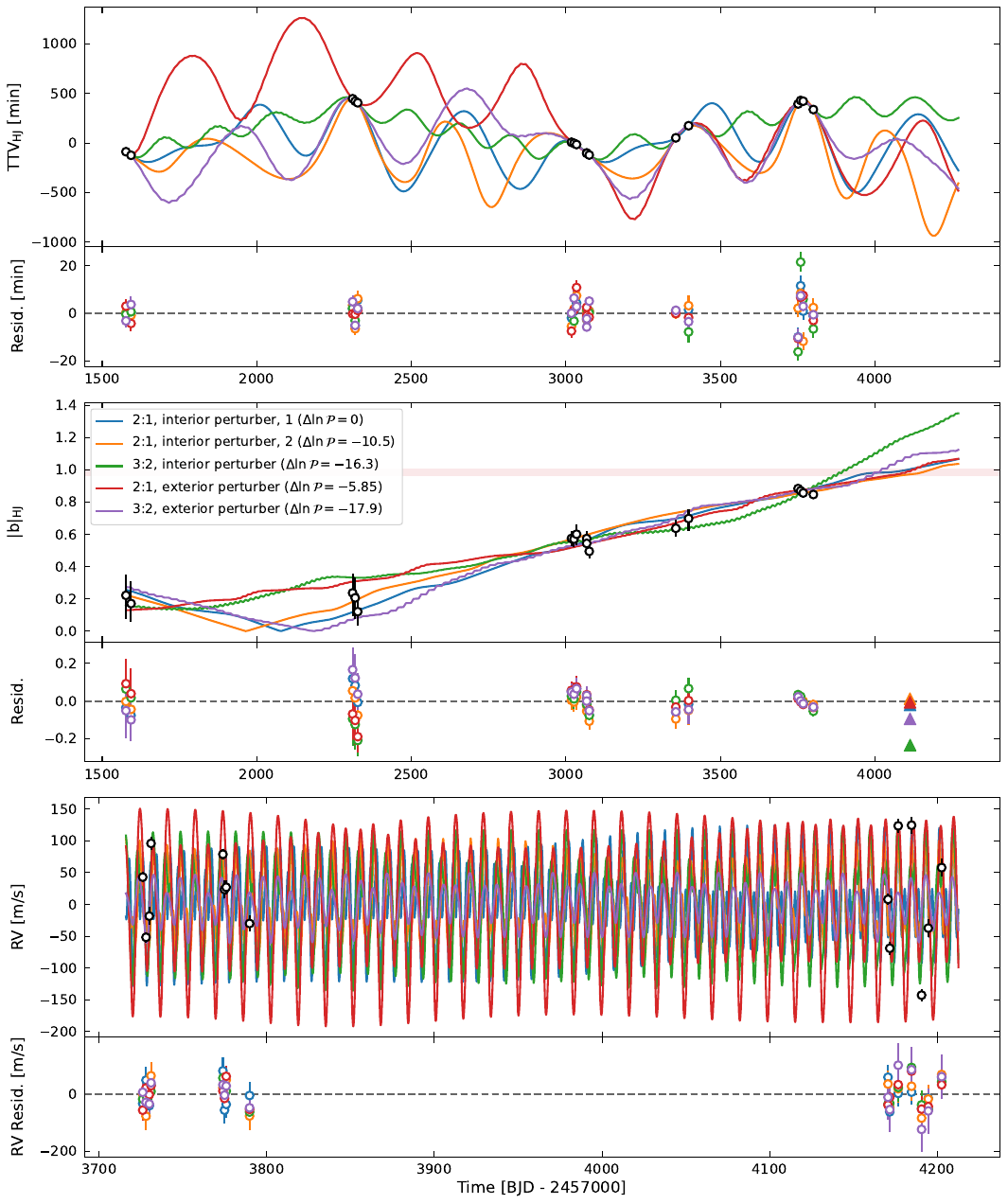}
\caption{Predicted transit times, impact parameters, and radial velocities for representative system architectures close to the 2:1 and 3:2 period commensurabilities. All of the plotted models produce reasonably good fits to the data, with log posterior probabilities within 20 of each other. The colored points in the residual panels show the residuals with respect to the corresponding model; with the RV residuals having error bars inflated by adding the best-fit jitter term in quadrature.
In the middle panel, the red shaded region denotes the non-detection constraint from Sector 101.
\label{fig:ttv_models}}
\end{figure*}

We investigated this preferred, two-planet architecture in greater detail by modeling the TTVs and TDVs with $N$-body simulations.
\citet{Lammers2026} found that it is typically not possible to uniquely identify unseen planets based on the TTVs and TDVs they induce on the observed transiting planets, even for cases with large TTV amplitudes.
Our observations of TOI-2535 are also sparsely sampled, making the problem of TTV inversion even more challenging.
Still, we sought to qualitatively understand what types of system architectures could explain the observations.

We began with $3\times10^5$ initial conditions, chosen as described in the following paragraph, and modeled the transit variations and RVs using the \texttt{TTVFast} $N$-body code \citep{TTVFast_Deck2014}.
These were treated as starting points for a local optimization, where we maximized the objective function
\begin{equation}
p(\theta\,|\,D) \propto \pi(\theta) \mathcal{L}_\mathrm{total}(D\,|\,\theta),
\end{equation}
where the total likelihood function is
\begin{equation}
\mathcal{L}_\mathrm{total} = \mathcal{L}_\mathrm{TTV} \times \mathcal{L}_\mathrm{TDV} \times
\mathcal{L}_\mathrm{RV}.
\end{equation}
As in Section \ref{ssec:nodal_precession}, we use a standard Gaussian likelihood for the TTVs and the empirical likelihood defined in Equation \ref{eq:tdv_likelihood} for $\mathcal{L}_\mathrm{TDV}$.
For the RV data, we assumed a Gaussian likelihood with a jitter term to account for underestimated uncertainties \citep[e.g.,][]{Howard2014}:
\begin{align}
\begin{split}
\mathcal{L}_\mathrm{RV} \propto \prod_{i=1}^{N}
&\frac{1}{\sqrt{2\pi\left(\sigma_i^2+\sigma_{\mathrm{jit},j(i)}^2\right)}} \times \\
&\exp\left[
-\frac{\left(v_i-v_{\mathrm{model}}(t_i) - \gamma_{j(i)} \right)^2}
{2\left(\sigma_i^2+\sigma_{\mathrm{jit},j(i)}^2\right)}
\right],
\end{split}
\end{align}
where $v_\mathrm{model}(t_i)$ is the predicted RV from \TTVFast at the observation times, while $\gamma_{j(i)}$ and $\sigma_{\mathrm{jit},j(i)}$ are an arbitrary offset and jitter term for instrument $j$ used for observation $i$.
We used uniform priors on each planet's initial orbital elements, and a log-uniform prior on their masses.

To choose the initial conditions used for the starting points of each local optimization, we used the following procedure.
One planet was chosen to have an initial orbital period, mean anomaly, and sky-plane inclination matching the earliest observations of TOI-2535\,b.
For the second planet, we scanned through a large range of period ratios, focusing on those surrounding first- and second-order mean-motion resonances (MMRs), as well as higher order MMRs of the form $J:1$. We also performed a coarser exploration of non-resonant period ratios ranging from $P_{\mathrm{HJ}}/P_2 = 1/50$ to $10/1$.
The other planetary parameters were initialized randomly, with an upper limit of 0.3 on the planets' initial eccentricities for the resonant cases and 0.8 for the non-resonant cases, and an upper limit of 30$^\circ$ on their mutual inclinations.
During the optimization, these parameters were allowed to vary beyond those bounds.

Figure \ref{fig:ttv_search} shows the results of this grid search and local optimization procedure.
We found that the current transit and RV data do not uniquely identify a system architecture, although hierarchical configurations are heavily disfavored, consistent with our arguments based on the nodal precession rate in \S\ref{ssec:nodal_precession}.
The posterior probability landscape shows deep and narrow valleys of good solutions clustered around MMRs.
The nominal maximum a posteriori (MAP) solution places the unseen perturber interior to the hot Jupiter, close to the 2:1 period commensurability, although in that particular solution the perturber is transiting and can be excluded.
Other nearby solutions, as well as those close to the 3:2, 3:1, and 4:1 period commensurabilities, also provide reasonable fits to the data and remain viable ($\Delta\ln p < 20$ compared with the nominal MAP solution).
We stress that our search is not exhaustive and even better solutions may be found with a deeper search.
This non-uniqueness is not surprising given the sparseness of our TTV and TDV data, limiting our ability to detect any high-frequency ``chopping'' TTV signal, which \citet{Lammers2026} found was typically necessary for breaking degeneracies between solution families.

We explored the 100 best solutions in greater detail by carrying out longer duration integrations with the \texttt{WHFast} integrator in \texttt{rebound} \citep{rebound_Rein2012,reboundwhfast_Rein2015b}.
We integrated each of these initial conditions for 1 Myr (roughly $5\times10^7$ orbits of the hot Jupiter), finding that approximately 2/3 of such systems underwent a close approach and instability within this time span.
None of the cases we explored that were close to second- and higher-order resonances turned out to be stable over the long-term.
Furthermore, we tracked the resonant dynamics during these integrations and found that in all stable cases, the resonant angles for eccentricity-type MMRs (e.g. $\psi = 2\lambda_\mathrm{out} - \lambda_\mathrm{in} - \hat{\varpi}$ for the 2:1 resonance case) were in libration.
We did not find any solutions that were actively in inclination-type resonances, although we cannot rule out that the system may have been in such a configuration in the past.

These considerations lead us to favor system architectures where the two planets are either in a 2:1 or 3:2 eccentricity-type MMR, in agreement with the general prevalence of these resonances in the broader exoplanet population.
Figure \ref{fig:ttv_models} and Appendix \ref{app:ttv_model_details} report planet parameters from stable solutions in these architectures.
While the perturber masses, eccentricities, and mutual inclinations vary between these solutions, we find that they generally favor a giant planet with $\Mp \gtrsim 0.3\,\Mjup$, as a more massive perturber is better able to drive the large TTVs and TDVs.

While our grid search procedure found that architectures with high mutual inclinations, up to $\imut \approx 60^\circ$ were able to explain the TTV, TDV, and RV data during the short-term integrations, these high-inclination solutions were far more likely to be unstable over the longer term.
Such systems typically saw the planets undergo chaotic evolution trading their high inclination for eccentricity, leading to eccentricity growth and subsequent close approach and instability.
Still, some configurations with moderate mutual inclinations proved to be stable over the 1 Myr integration time, such as the ``3:2, interior perturber'' case reported in Table \ref{tab:ttv_model_params}, where the two planets have a mutual inclination of $\imut = 33^\circ$.

Tracking the evolution of the orbital elements from our $N$-body integrations reveals that the main mechanism driving the impact parameter change is indeed nodal precession at a constant $\dot{\Omega}_\mathrm{inv}$, with precession timescales $P_\Omega$ of 30--50 years.
Note that for a two-planet system, the planets' orbits have $\Omega_\mathrm{1,inv} = \Omega_\mathrm{2,inv} + \pi$ and precess at the same rate.
The deviations in the model predictions shown in Figure \ref{fig:ttv_models} from a straight line are due to variations in the eccentricity and apsidal angle arising from resonant interactions between the planets, contributing a secondary effect to the variation in $b$.
We did not find any system architectures with $i_\mathrm{mut} < 5^\circ$ that could cause a sufficiently large precession rate to explain the TDVs.
If this is indeed the minimum required mutual inclination, TOI-2535 would be one of just a handful of compact multi-planet systems known to have significant mutual inclinations between planets.

The best-fit models predict TTV curves that are not simple sinusoidal curves, but have two or more dominant frequencies, as expected from our preliminary analysis in Section \ref{sec:light_curve_analysis}. For systems in MMRs, TTVs are primarily due to to libration of the resonant angle; we find that this is the case from our integrations (e.g. Fig. \ref{fig:candidate_029_resonance}), which show this angle librating at multiple frequencies.
One caution as to the interpretation of these results is that the timing data show significant excess scatter even for the best-fitting models, with residuals of up to 10 minutes, several times the nominal timing uncertainties.
This could be due to unresolved ``chopping'' signals, or an underestimation of the uncertainties in the timing measurements.
The two-planet models also provided only a modest improvement to the fit to the RV data compared with a single-planet Keplerian model (Fig. \ref{fig:rv_periodogram}), bringing the required RV jitter down from 55 to $30$~\ms, still larger than expected for an apparently old and quiet K star \citep[e.g.,][]{Luhn2020}.
The large remaining RV and timing jitter could point to the possibility that there is more than one unseen planet in the system.
Since continued transit timing monitoring is not possible, a high-cadence RV campaign is the best path forward to unveiling the true architecture of the system.
The relatively large masses for the perturber found by our best-fit TTV models imply RV semi-amplitudes of 30--120~\ms, which should in principle allow it (and any other bodies in the system) to be independently detected by a larger and more finely sampled RV dataset.

\section{Discussion \& Summary}

The TOI-2535 system is an unusual and exceptional system in many ways. Observationally, it has the largest TDVs and inclination change ever observed for a planet orbiting a single star, with a precession period of decades, shorter than other known systems (e.g. the TOI-4504 system has $P_\Omega = 250$~years, \citealp{TOI4504_Almenara2026}; while TOI-216 has $P_\Omega = 150$~years, \citealp{TOI216_McKee2023}).
The TTV semi-amplitude of 6 hours, 3\% of the planet's orbital period, is also among the largest TTV signals known, both in absolute and fractional terms (see e.g. Figure 10 of \citealp{TOI4504_Vitkova2025}), and is certainly the largest for a hot Jupiter with $\Porb < 10$~days.

Even though the nature of the perturber to TOI-2535\,b could not be uniquely pinned down, the large TTVs it drives require it to be in or near a mean-motion resonance.
This makes TOI-2535\,b one of only a handful of hot Jupiters known to have a close-by planetary companion \citep{Sha2026}.
Furthermore, the companion to TOI-2535\,b is likely a second short-period giant planet. This differs from all but one of the currently known hot Jupiters with a nearby companion, in which the companions are sub-Neptunes or super-Earths.\footnote{The exception is the WASP-148 system \citep{WASP148_Hebrard2020}, which contains a hot Jupiter and an outer giant planet near a 4:1 period commensurability.}
Such systems could not have formed through high-eccentricity migration \citep[e.g.,][]{Fabrycky2007}, which has been posited to be responsible for most other hot Jupiters.
Instead, the two planets in the TOI-2535 system likely underwent migration through torques with the protoplanetary disk, a process that naturally explains their resonant capture into an MMR \citep{Baruteau2014}.

However, the protoplanetary disk would also quickly damp the inclinations of the two planets \citep{Tanaka2004}, requiring some later excitation to explain the non-zero inclination required by the TDVs.
\citet{Thommes2003} showed that the inclination of two migrating planets could be excited by tens of degrees if they passed through an inclination-type resonance.
Alternatively, a sweeping secular resonance, either from an evolving protoplanetary disk \citep{Petrovich2020,Zanazzi2024} or stellar spin-down \citep{Spalding2016,Faridani2024}, could also excite mutual inclinations.
These mechanisms can only operate in specific regions of the space of possible planetary orbital parameters, so pinning down the true architecture of this system could reveal whether either of these is responsible for the system's current configuration.

Identifying the perturber mass and location would provide a secure constraint on the actual mutual inclination in the system.
If the $\imut > 5^\circ$ implied by our TTV analysis is confirmed, TOI-2535 would be one of only a few compact multi-planet systems known to have such large mutual inclinations between planets.
Most systems with multiple transiting planets in compact architectures have low mutual inclinations of 1--2$^\circ$ \citep{Fabrycky2014}, and even the other known cases with large TDVs have similarly low mutual inclinations (\citet{TOI4504_Almenara2026} reported $\imut = 2.1^\circ$ between TOI-4504\,c and d, while \citet{TOI216_McKee2023} found $\imut = 1.5^\circ$ for TOI-216\,b and c).
Recently, \citet{Quinn2026} examined the sample of hot Jupiters with nearby transiting companions and found preliminary evidence that their typical mutual inclinations are low, similar to the compact systems of smaller planets found by \Kepler.

Of course, TOI-2535 was detected as a system with only a single transiting planet, so statistics based on systems containing multiple transiting planets may not be applicable.
\citet{He2020} modeled the intrinsic multiplicity and inclination distribution of \Kepler-discovered systems, including undetected non-transiting planets, and found that the data are consistent with a population of systems that have mutual inclinations that depend on their intrinsic multiplicity, such that two-planet systems have a median mutual inclination of $5.4^\circ$.
\citet{Millholland2021b} later corroborated this interpretation using the distribution of TDVs in the \Kepler sample.
Thus, despite the relatively larger masses of the planets in the TOI-2535 system compared with the super-Earths and sub-Neptunes that comprise the majority of the planets found by \Kepler, if they are in one of the low mutual inclination configurations found in \S\ref{ssec:dynamical_analysis}, they could be consistent with being part of this broader population.

TESS' all-sky transit survey of millions of stars enables it to find relatively rare objects, while its increasing observation baseline, now approaching double that of the \Kepler mission, enables sensitivity to long-term transit variations.
Indeed, given just the first four years of TESS data, the TDVs of the TOI-2535 system would not have been securely detected, and it is only with continued monitoring that it was revealed to have some of the most extreme TDVs known.
As TESS continues to monitor the sky, more systems with TDVs could be revealed, and these will be a key sample for understanding mutual inclinations in exoplanetary systems and the processes that excite them.

\begin{acknowledgments}
S.W.Y. gratefully acknowledges support from the Heising-Simons Foundation. The authors thank the observers at the WIYN 3.5m telescope for their skillful execution of the observations, and the NEID instrument team for providing the reduced data products.
S.W.Y. thanks Jiayin Dong for helpful discussions.
R.B.\ acknowledges support from FONDECYT Project 1241963.
A.J.\ acknowledges support from FONDECYT project 1251439.
Funding for K.B. was provided by the European Union (ERC AdG SUBSTELLAR, GA 101054354).
F.M. and E.P. acknowledge financial support from the Agencia Estatal de Investigaci\'{o}n (AEI) through the Severo Ochoa Centre of Excellence accreditation awarded to the Instituto de Astrof\'{i}sica de Canarias, grant CEX2025-001609-S, funded by MICIU/AEI/10.13039/501100011033.

This paper includes data collected by the TESS mission, which are publicly available from the Mikulski Archive for Space Telescopes (MAST). Funding for the TESS mission is provided by the NASA's Science Mission Directorate. We acknowledge the use of public TOI Release data from pipelines at the TESS Science Office and at the TESS Science Processing Operations Center. This research has made use of the Exoplanet Follow-up Observation Program website, which is operated by the California Institute of Technology, under contract with the National Aeronautics and Space Administration under the Exoplanet Exploration Program.

Part of the data presented herein were obtained with the WIYN 3.5m/NEID from telescope time allocated under the NN-EXPLORE program, with support from the National Aeronautics and Space Administration and the U.S. National Science Foundation. Some of the data were obtained with the NEID spectrograph built by Penn State University and operated at the WIYN Observatory by NSF NOIRLab, under the NN-EXPLORE partnership of the National Aeronautics and Space Administration and the U.S. National Science Foundation. The authors are honored to be permitted to conduct astronomical research on I'oligam Du'ag (Kitt Peak), a mountain with particular significance to the Tohono O'odham.
This work makes use of observations from the Las Cumbres Observatory global telescope network.

\end{acknowledgments}

\facilities{TESS, WIYN (NEID), FLWO:1.5m (TRES), LCOGT}

\software{\texttt{astropy} \citep{Astropy13,Astropy18,Astropy2022},
          \texttt{emcee} \citep{Emcee_Foreman-Mackey13,Emcee_Goodman10}
          \texttt{EXOFASTv2} \citep{ExoFAST_Eastman2013,ExoFASTv2_Eastman19},
          \texttt{exoplanet} \citep{exoplanet_ForemanMackey2021,exoplanet_Agol2020},
          \texttt{lightkurve} \citep{Lightkurve18},
          \texttt{rebound} \citep{rebound_Rein2012},
          \texttt{SpecMatch-Emp} \citep{SpecMatchEmp_Yee2017},
          \texttt{TTVFast} \citep{TTVFast_Deck2014}.}

\bibliography{paper,software,instruments,catalogs}{}
\bibliographystyle{aasjournalv7}

\restartappendixnumbering
\appendix
\section{Confirmation of TOI-2535 b from Ground-Based Observations}
\label{app:planet_validation}

In this Appendix, we describe how the ground-based follow-up observations eliminate potential false positive scenarios and confirm TOI-2535\,b as a \textit{bona fide} planet.

Given the $21^{\prime\prime}$ pixel scale of TESS, a common false positive scenario occurs when the deep eclipses from a nearby stellar eclipsing binary are blended with the light of a bright star, resulting in apparent shallow flux decrements that can be mistaken with planetary transits.
The time-series photometry obtained from the ground (Table \ref{tab:sg1_summary}) are seeing-limited and resolve all known stellar sources from the \Gaia DR3 catalog, the closest of which is separated by $18^{\prime\prime}$. 
These observations thus confirm that the transit is on-target and not due to deep eclipses on neighboring faint sources.
Furthermore, the observations from the LCO SAAO node on 2023 May 02 were taken in alternating $g^\prime$ and $i^\prime$ filters, where we found consistent transit depths in these two bands for the same transit event, whereas eclipsing binary false positive scenarios due to stars of different temperatures to the primary produce chromatic transit events.

\begin{figure}
    \centering
    \includegraphics[width=\linewidth]{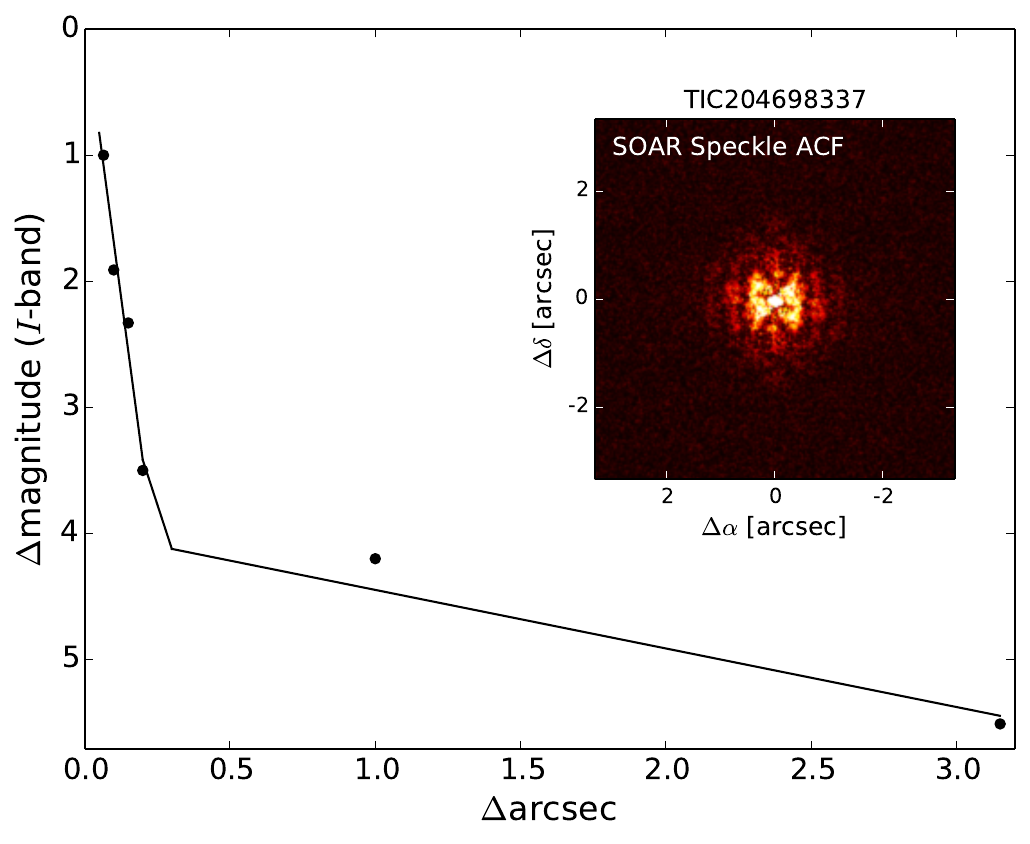}
    \caption{SOAR HRCam observation of TOI-2535. The black line shows the detection limits as a function of separation, while the inset image shows the speckle autocorrelation function (ACF).}
    \label{fig:soar_imaging}
\end{figure}

Speckle imaging observations from SOAR HRCam allowed us to check for the presence of stellar binaries even closer than the $\approx1^{\prime\prime}$ seeing limit. 
As Figure \ref{fig:soar_imaging} shows, no bright companions were detected down to the instrumental detection limits.
This is corroborated by the lack of astrometric excess noise as measured by \Gaia, as the reported RUWE of 1.12 is lower than the expected threshold of $\sim 1.22$ that a binary system is expected to produce given the spacecraft scanning law and sky position of TOI-2535 \citep{GaiaUnlimited_Castro-Ginard2024}.

The spectroscopic observations provided the final data confirming the planetary nature of TOI-2535\,b.
These observations confirmed that TOI-2535 has a single-lined spectrum, with no evidence for spectroscopic binarity.
Meanwhile, no significant radial-velocity (RV) shifts were detected between the two FEROS observations within their uncertainties, ruling out the possibility that the transits are due to a brown dwarf or stellar mass object, since a massive companion would cause RV variations at the \kms level.

\begin{figure}
    \centering
    \includegraphics[width=\linewidth]{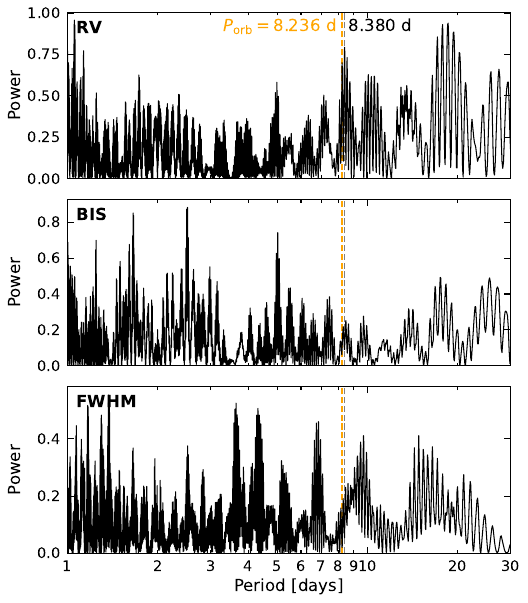}
    \includegraphics[width=\linewidth]{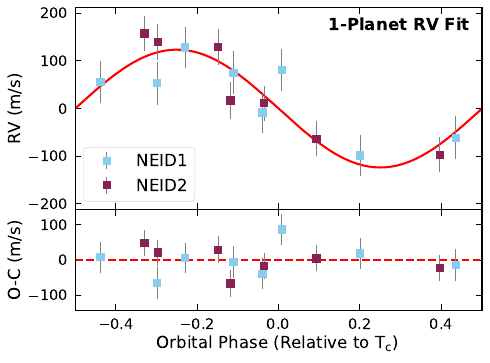}
    \caption{Top: Lomb-Scargle periodograms of the NEID RV, BIS, and FWHM measurements.
    There is a cluster of peaks close to $\Porb = 8.236$~days, with the fine structure due to aliasing with a 432~day peak in the window function representing the characteristic spacing between the two observing seasons.
    Bottom: Phase-folded RV data (squares) and best-fit model (red line) from a single Keplerian fit to the RV data, with $P$ held fixed to $\Porb$ and $T_\mathrm{c}$ allowed to vary within $\pm 6$ hours of the best-fit linear ephemeris to the transit data. The error bars shown here include both the instrumental uncertainties and a fitted jitter value, added in quadrature.}
    \label{fig:rv_periodogram}
\end{figure}

The NEID RV data are reported in Table \ref{tab:rv_obs}, and a periodogram of these measurements (Figure \ref{fig:rv_periodogram}) reveals a cluster of peaks close to the linear ephemeris implied by the transit data, indicating a potential periodic signal at this period, although there is some aliasing due to the uneven and sparse time sampling of the RV observations.
As part of a preliminary analysis of these data, we fitted them with a Keplerian model for a single planet on a circular orbit, with orbital period fixed to the best-fit linear ephemeris from the transit data, while the time of conjunction was allowed to vary within $\pm 6$~hours from the transit ephemeris.
Because of a failure in NEID's temperature control system that led to an RV offset between observing seasons, we treat the data taken in 2025 and 2026 as coming from separate instruments with independent offsets and RV jitter terms.
The lower panel of Figure \ref{fig:rv_periodogram} shows that the RV data are consistent with the presence of a massive planet orbiting at the transit ephemeris, with $K = 108^{+32}_{-39}\,\ms$, corresponding to $M_p\sin{i} = 0.99^{+0.30}_{-0.33}\,\Mjup$.
However, the TTVs and TDVs indicate that there is at least one other planet in the system, so this single-planet fit is only illustrative; indeed, the large RV jitter required to bring the reduced $\chi^2$ to 1 ($\jit \sim 55~\ms$) likely points to additional unmodeled signals.

The NEID-DRP also produces bisector inverse slope (BIS) and full-width at half-maximum (FWHM) measurements of the combined CCF.
No significant power is seen at the transit orbital period in periodograms of the BIS and FWHM measurements.
We also saw no correlation between the RVs and the BIS and FWHM measurements.
These tests indicate that the RV variation is likely due to the influence of a transiting planet, rather than distortions in the stellar line shape due to stellar activity or an unresolved eclipsing binary.

In summary, our ground-based follow-up observations provide no indications that the transits detected by TESS could be due to a resolved or unresolved stellar binary, while the NEID RV variations rule out a stellar-mass companion on a close-in orbit around TOI-2535.
We therefore consider TOI-2535\,b to be a confirmed exoplanet, even though the degeneracies in the TTV modeling preclude a unique mass measurement.

\begin{deluxetable}{ccc} \label{tab:rv_obs}
\tablecaption{NEID RV Measurements}
\tablehead{
\colhead{Obs. Midpoint} & \colhead{Relative RV} & \colhead{$\sigma_\mathrm{RV}$}  \\
\colhead{\bjdtdb} & \colhead{[\ms]} & \colhead{[\ms]} 
}
\startdata
2460725.951524 & 31262.8 & 8.2 \\
2460727.927528 & 31177.0 & 8.2 \\
2460729.868637 & 31219.0 & 14.0 \\
2460730.907921 & 31338.0 & 10.3 \\
2460773.796976 & 31512.7 & 8.2 \\
2460774.773015 & 31461.8 & 14.6 \\
2460775.756529 & 31469.6 & 11.7 \\
2460789.698983 & 31475.0 & 12.8 \\
2461170.674968 & 31224.5 & 9.2 \\
2461171.732000 & 31152.3 & 10.6 \\
2461176.755580 & 31367.3 & 11.2 \\
2461184.723412 & 31403.9 & 13.3 \\
2461190.697334 & 31163.4 & 9.4 \\
2461194.701001 & 31286.7 & 14.8 \\
2461202.684039 & 31417.6 & 14.7%

\enddata
\end{deluxetable}

\begin{rotatepage}
\movetabledown=2in
\begin{rotatetable}
\begin{deluxetable*}{ccccccccc}
\tabletypesize{\footnotesize}
\tablecolumns{9}
\tablecaption{Summary of Ground-Based Transit Observations of TOI-2535\,b
\label{tab:sg1_summary}}
\tablehead{
    \colhead{Epoch\tablenotemark{a}} &
    \colhead{Facility/Instrument} & \colhead{Location} & \colhead{Aperture} & \colhead{Filter} &
    \colhead{Date} & \colhead{Cadence} & \colhead{Transit Detected} & \colhead{Detrending Vectors\tablenotemark{a}} \\
    & & & \colhead{(m)} & & \colhead{(UT)} & \colhead{(s)} & &
}
\startdata
-- & El Sauce & Coquimbo, Chile & 0.51 & $R$ & 2021 Apr 12 & 180 & N & -- \\
-- & LCO CTIO/Sinistro & La Serena, Chile & 1.0 & $g^\prime$ & 2021 Apr 28 & 120 & N & -- \\
-- & LCO CTIO/Sinistro & La Serena, Chile & 1.0 & $i^\prime$ & 2021 Apr 28 & 70 & N & -- \\
-- & TRAPPIST-South & La Silla, Chile & 0.6 & $i^\prime$ & 2021 Apr 28 & 35 & N & -- \\
-- & Brierfield & Bowral, NSW, Australia & 0.36 & $I$ & 2021 May 15 & 180 & N & -- \\
-- & Brierfield & Bowral, NSW, Australia & 0.36 & $R$ & 2023 Apr 21 & 240 & N & -- \\
181 & LCO SAAO/Sinistro & Sutherland, South Africa & 1.0 & $g^\prime$ & 2023 May 02 & 120 & Y & Airmass, Width \\
181 & LCO SAAO/Sinistro & Sutherland, South Africa & 1.0 & $i^\prime$ & 2023 May 02 & 70 & Y & Total Counts, Width \\
182 & LCO CTIO/Sinistro & La Serena, Chile & 1.0 & $i^\prime$ & 2023 May 10 & 70 & Y & J.D.-2400000, Airmass, Width, Sky/Pixel \\
216 & LCO Teide/SBIG-6303 & La Laguna, Tenerife, Spain & 0.4 & $g^\prime$ & 2024 Feb 15 & 300 & Y & J.D.-2400000 \\
-- & LCO CTIO/SBIG-6303 & La Serena, Chile & 0.4 & $g^\prime$ & 2024 Mar 27 & 300 & N & -- \\
221 & LCO Haleakala/SBIG-6303 & Maui, HI, USA & 0.4 & $g^\prime$ & 2024 Mar 27 & 300 & Y & BJD$_\mathrm{TDB}$ \\
-- & LCO McDonald/SBIG-6303 & Fort Davis, TX, USA & 0.4 & $g^\prime$ & 2024 Mar 27 & 300 & N & -- \\
232 & LCO CTIO/SBIG-6303 & La Serena, Chile & 0.4 & $i^\prime$ & 2024 Jun 25 & 300 & Y & Width \\
270 & LCO CTIO/SBIG-6303 & La Serena, Chile & 0.4 & $i^\prime$ & 2025 May 04 & 300 & Y & BJD$_\mathrm{TDB}$ \\
-- & SUTO/OTIVAR & Pyskowice, Poland & 0.3 & $R$ & 2026 Mar 13 & 240 & N & -- \\
-- & LCO McDonald/SBIG-6303 & Fort Davis, TX, USA & 0.4 & $g^\prime$ & 2026 Mar 22 & 200 & N & -- \\
-- & LCO McDonald/SBIG-6303 & Fort Davis, TX, USA & 0.4 & $i^\prime$ & 2026 Mar 22 & 200 & N & --%
\enddata
\tablenotetext{a}{Epoch numbers and detrending vectors are only provided for observations where a transit was securely detected in the data.}
\end{deluxetable*}
\end{rotatetable}
\end{rotatepage}

\section{Stellar Characterization}
\label{app:stellar_characterization}

We characterized the host star, TOI-2535, using spectroscopic, photometric, and astrometric measurements.
First, we applied the \texttt{SpecMatch-Empirical} code \citep{SpecMatchEmp_Yee2017} to measure stellar atmospheric properties from the NEID spectra.
We performed a blaze and barycentric correction to the individual NEID observations before co-adding them to obtain a combined stellar spectrum with a higher signal-to-noise ratio.
\texttt{SpecMatch-Empirical} then compares the target spectrum to a library of observed spectra from stars with well-measured properties, and reports parameters interpolated from the best-matching library stars.
Table \ref{tab:stellar_params} reports the inferred atmospheric properties for TOI-2535.

We then used the \texttt{EXOFASTv2} code \citep{ExoFASTv2_Eastman19} to fit the observed stellar spectral energy distribution (SED) to the bolometric correction grids derived for the MIST stellar evolutionary models \citep{MIST0_Dotter2016,MISTI_Choi2016}.
We fitted the broadband photometry from the \Gaia, 2MASS, and WISE surveys, imposing error floors of 0.02~mag on the \Gaia photometry. We placed a Gaussian prior on the stellar metallicity based on the spectroscopic measurement from \texttt{SpecMatch-Emp}, as well as an upper limit on the visual extinction $A_V$ based on the dust maps of \citet{Schlegel1998} and \citet{Schlafly2011}.
The results from the SED fit agree with the spectroscopic analysis, and we find that TOI-2535 is an old, early K dwarf with $\Mstar = 0.86\pm0.03\,\Msun$.

\begin{figure}
    \centering
    \includegraphics[width=\linewidth]{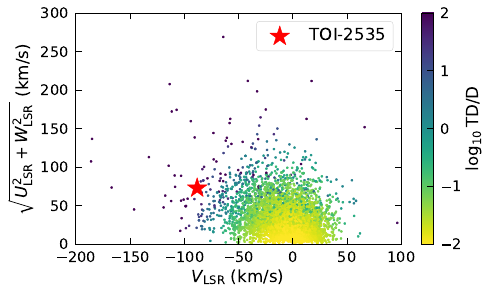}
    \caption{Toomre diagram showing the space velocity of TOI-2535 relative to other TESS Objects of Interest, as computed from their \Gaia astrometry. Points are color-coded with the odds ratio of membership in the thick or thin disk (TD/D) based on the methods of \citet{Bensby2014}; the high space velocity of TOI-2535 suggests it may be a member of the galactic thick disk.}
    \label{fig:toomre}
\end{figure}

We also investigated the 6D galactic kinematics of TOI-2535 using astrometric and absolute RV measurements from \Gaia DR3 \citep{GaiaDR3_Vallenari2022}.
Intriguingly, TOI-2535 appears to have a moderately high space velocity relative to the local standard of rest, $V_\mathrm{tot, LSR} = 114$~\kms (Figure \ref{fig:toomre}).
Using the classifications from \citet{Bensby2014}, we find that TOI-2535 has kinematics consistent with that of a galactic thick disk star ($Pr(\mathrm{thick\,disk}) = 98.8\%$), with only a 1.2\% probability of being a thin disk star.
We also used the \texttt{gala} package \citep{Gala_PriceWhelan2017,Gala_PriceWhelan2020} to integrate the galactic orbit of TOI-2535 within the \texttt{MilkyWayPotential2022} \citep{Gala_Eilers2019}, finding that it is on a moderately eccentric ($e = 0.42$, $\eta \equiv L_z/L_c = 0.82$) and inclined ($z_\mathrm{max} = 0.39$~kpc) orbit, further evidence for membership in the Milky Way's kinematically hot thick disk \citep[e.g.,][]{Chandra2024}.
The old age inferred from the MIST model is also consistent with this galactic population classification, although definitive conclusions will require a measurement of the host star's $\alpha$-element abundance.

\section{Transit Fit Parameters \label{app:transit_fit_params}}

Table \ref{tab:transit_params} reports the fit parameters and priors from the light curve fitting described in Section \ref{sec:light_curve_analysis}, and we explain our prior choices here.
We used a Gaussian prior of $\mathcal{N}(1.8, 0.2)$ in units of \gcc on \rhostar based on the stellar characterization described in Appendix \ref{app:stellar_characterization}.
The parameterization of eccentricity in terms of (\secosw, \sesinw) imposes an implicit uniform prior on $e$ \citep{Anderson2011} while avoiding a hard boundary at $e = 0$.

For each transit, we fitted separate $b$ and $T_c$ values. 
We placed a uniform 0.2-day prior on each $T_{c,i}$ around an approximate mid-transit time from a visual inspection of the photometry.
As mentioned in Section \ref{sec:light_curve_analysis}, 
we fitted for a signed value of $b$ for the TESS transits from Sectors 10 and 37. For these data, we used the uniform prior $b_i\sim \mathcal{U}(-1.0, 1.0)$, while for the remaining transits, their inferred impact parameters are well-separated from zero and we restricted their $b_i$ to be positive with the prior $b_i \sim \mathcal{U}(0.0, 1.0)$.
We note that \citet{Dawson2020} argued against the use of uniform priors on $b$ when modeling TDVs, instead advocating for a hierarchical model with a prior on the mean impact parameter and scale of impact parameter change.
In our tests, we found that we were able to avoid the biases found by \citet{Dawson2020} by using the full seven-parameter basis with a physically motivated prior on $\rho_\star$, rather than a five-parameter basis that includes the stellar pseudodensity $\rho_\mathrm{circ}$.

We assumed a quadratic limb-darkening law in the transit model, fitting for the limb-darkening coefficients in terms of the $(q_1,\,q_2)$ basis specified by \citet{Kipping2013}.
Transits observed in the same filter shared the same limb-darkening coefficients.
Each transit model was computed at 1 minute cadence and then integrated across the actual observation cadence for each dataset, to correctly account for the smearing of the transit shape, especially in the long-cadence TESS data.
\begin{deluxetable}{ccc}
\tablecolumns{3}
\tablecaption{Joint Transit Fit Parameters \label{tab:transit_params}}
\tablehead{
    \colhead{Parameter} & \colhead{Prior} & \colhead{Posterior}
}
\startdata
$P_\mathrm{orb,lin}$ (days)\tablenotemark{a} & Derived & $8.2358054 \pm 0.0000079$ \\
$T_{0,\mathrm{lin}}$ ($\bjdtdb - 2457000$)\tablenotemark{a} & Derived & $1576.8351^{+0.0010}_{-0.0013}$ \\
$R_p/R_\star$ & $\mathcal{U}(0.01, 0.5)$ & $0.0989^{+0.0013}_{-0.0011}$ \\
\rhostar (\gcc) & $\mathcal{N}(1.8, 0.2)$ & $1.84^{+0.20}_{-0.21}$ \\
\secosw & $\mathcal{U}(-0.9, 0.9)$ & $-0.03^{+0.52}_{-0.53}$ \\
\sesinw & $\mathcal{U}(-0.9, 0.9)$ & $0.13^{+0.19}_{-0.13}$ \\
\ecosw & Derived & $-0.01^{+0.29}_{-0.27}$ \\
\esinw & Derived & $0.049^{+0.075}_{-0.052}$ \\
$e$ & Derived & $0.19^{+0.13}_{-0.16}$ \\
$\omega$ (deg) & Derived & $55^{+120}_{-45}$ \\
$a/R_\star$ & Derived & $18.74^{+0.68}_{-0.70}$ \\
\Rp (\Rjup) & Derived & $0.844^{+0.027}_{-0.029}$
\enddata
\tablenotetext{a}{These ephemerides are computed by taking a linear least squares fit to the individual transit times reported in Table \ref{tab:transit_times}.}
\end{deluxetable}

\section{Representative TTV Models}
\label{app:ttv_model_details}

In this Appendix, we describe our $N$-body explorations of some representative system architectures in more detail.
Table \ref{tab:ttv_model_params} provides the model parameters corresponding to five solutions spanning four first-order MMRs; the TTV, TDV, and RV predictions from these cases are plotted in Figure \ref{fig:ttv_models}.

\begin{figure}
    \centering
    \includegraphics[width=\linewidth]{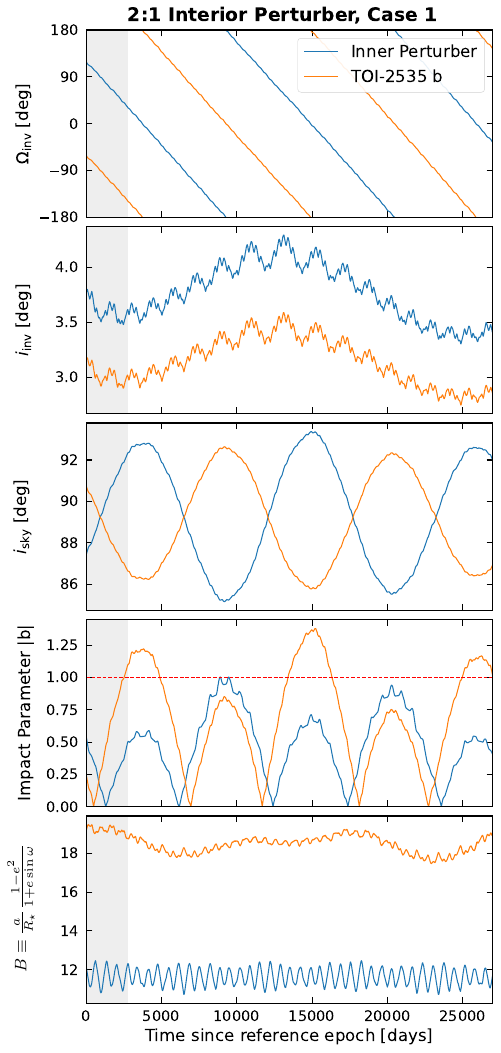}
    \caption{Variation in orbital elements for the ``2:1, interior perturber case 1'' solution. The top two panels show the longitude of ascending node and inclination, measured with respect to the system invariable plane. The third panel shows the inclination with respect to the sky plane, while the fourth panel shows the unsigned impact parameter. The fifth panel shows the variation in $B \equiv \frac{a}{R_\star} \frac{1-e^2}{1+e\sin{\omega}}$. The grey shaded region corresponds to the time spanned by the observed data and plotted in Figure \ref{fig:ttv_models}.}
    \label{fig:candidate_032_inclination}
\end{figure}

Across our grid search and local optimization procedure, the solution with the greatest posterior probability and that was long-term stable has a perturbing planet in an interior 2:1 MMR with TOI-2535\,b (``2:1, interior perturber, 1'' in Table \ref{tab:ttv_model_params}). Figure \ref{fig:candidate_032_inclination} shows the variation in orbital elements over 27{,}000 days, roughly 10$\times$ the duration spanned by the observations.
The top panel illustrates the regression of the $\Omega_\mathrm{inv}$ at a constant rate, as measured in the invariable plane, which results in an oscillating sky-plane inclination $i_\mathrm{sky}$.
We computed $\dot{\Omega}_\mathrm{inv}$ and hence $P_\Omega$ in Table \ref{tab:ttv_model_params} from a linear fit to the evolution of $\Omega_\mathrm{inv}$.
Resonant interactions do cause variations in $i_\mathrm{inv}$; this effect is smaller and barely affects the observed sky-plane inclination.
We also plot the variations in $B \equiv \frac{a}{R_\star} \frac{1-e^2}{1+e\sin{\omega}}$. We find that $B$ varies at the 10\% level, confirming the influence of apsidal precession and eccentricity variations on $b$ is subdominant compared with that caused by nodal precession, as discussed in \S\ref{ssec:nodal_precession}.

While Figure \ref{fig:candidate_032_inclination} is specific to the 2:1, interior perturber solution, the relative importance of the effects is similar across all the cases we explored.
Although this particular architecture provided a good fit to the TTVs and TDVs of TOI-2535\,b, it predicts that the interior perturber would have transited during the TESS observations. Given its inferred mass of $0.5\,\Mjup$, comparable to that of TOI-2535\,b, such transits would likely have been detected, and so we can exclude this as a plausible solution.
Other solutions in the 2:1 MMR remain viable, such as that provided in ``case 2'' listed in Table \ref{tab:ttv_model_params}.

\begin{figure}
    \centering
    \includegraphics[width=0.985\linewidth]{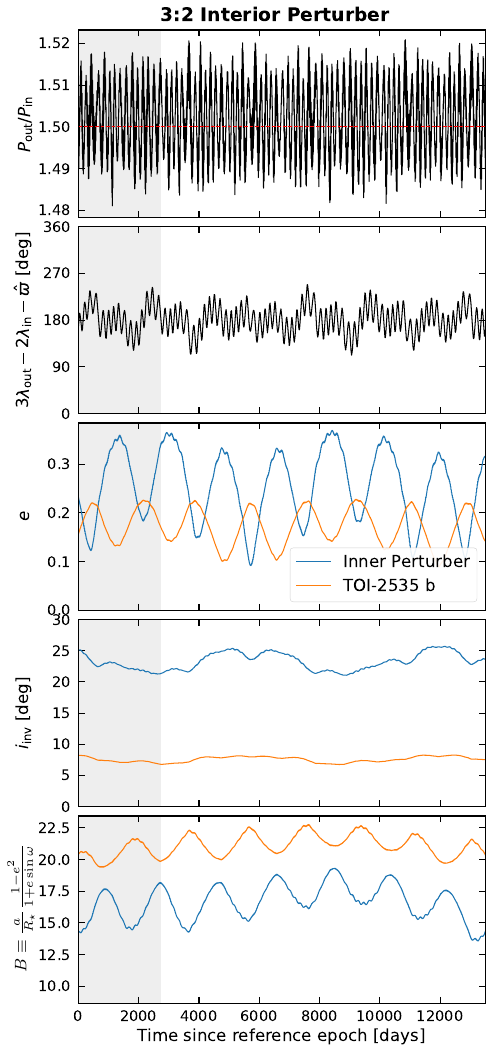}
    \caption{Variation in orbital elements for the ``3:2, interior perturber'' solution. The top panel shows the period ratio, while the second panel shows the mixed resonant angle. The two planets' orbital eccentricities and inclinations are shown in the third and fourth panels, while the fifth panel shows the parameter $B$.}
    \label{fig:candidate_029_resonance}
\end{figure}

We also explored the resonant behavior of these solutions. Figure \ref{fig:candidate_029_resonance} provides an illustrative case for the 3:2, interior perturber solution.
The mixed resonant angle, $\psi = 3\lambda_\mathrm{out} - 2\lambda_\mathrm{in} - \hat{\varpi}$, where $\hat{\varpi} = \mathrm{Arg}(fz_1 + gz_2)$ and $z_1,\,z_2$ are the complex eccentricities, librates around 180$^\circ$, as expected for a system in resonance. Both the eccentricities and inclinations of the two planets are moderate, so it is likely thanks to the phase protection provided by the MMR that this architecture is stable over a 1~Myr integration time.
As noted in the main text, among the solutions that fit the observations well, all stable cases had librating resonant angles.

\begin{figure}
    \centering
    \includegraphics[width=\linewidth]{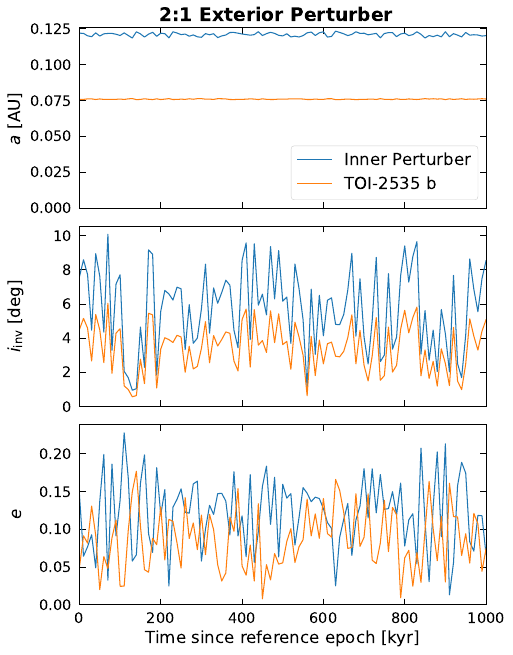}
    \caption{Long-term variation in orbital elements for the ``2:1, exterior perturber'' solution. Over 1 Myr, the inclinations and eccentricities of the two planets undergo chaotic evolution but remain bounded, with no close approaches during the integration time.}
    \label{fig:candidate_005_longterm}
\end{figure}

Finally, Figure \ref{fig:candidate_005_longterm} shows an example of the long-term evolution of orbital elements for the 2:1, exterior perturber solution.
The inclinations and eccentricities vary chaotically, but remained bounded.

\begin{deluxetable}{lccccc}
\tablecolumns{6}
\tablecaption{Planet Parameters for Representative Solutions from TTV Search \label{tab:ttv_model_params}}
\tablehead{
& \colhead{2:1} & \colhead{2:1} & \colhead{3:2} & \colhead{2:1} & \colhead{3:2} \\
& \colhead{interior perturber, 1} & \colhead{interior perturber, 2} & \colhead{interior perturber} & \colhead{exterior perturber} & \colhead{exterior perturber} 
}
\startdata
\multicolumn{6}{l}{\textbf{TOI-2535\,b}} \\
$P$ (d) & 8.2351108 & 8.2285457 & 8.2150137 & 8.200001 & 8.225629 \\
$e \cos \omega$ & 0.00854831 & 0.0849848 & -0.106131 & 0.0503934 & -0.0975226 \\
$e \sin \omega$ & -0.0331046 & -0.0490361 & -0.115584 & -0.0030093 & -0.0689933 \\
$i_\mathrm{sky}$ (deg) & 90.7677 & 89.3339 & 89.547 & 89.6054 & 90.7995 \\
$M$ (deg) & 160.157 & 105.516 & -128.547 & 83.3927 & -117.985 \\
$M_p$ ($M_{\rm Jup}$) & 0.472679 & 0.580814 & 0.85672 & 1.14837 & 0.352019 \\
\multicolumn{6}{l}{\textit{Planet 2}} \\
$P$ (d) & 4.0963862 & 4.1130895 & 5.5176076 & 16.764866 & 12.478175 \\
$e \cos \omega$ & -0.00697603 & 0.0706903 & -0.222522 & -0.104587 & -0.135774 \\
$e \sin \omega$ & -0.0477419 & -0.0549047 & -0.0640873 & -0.10204 & -0.0458248 \\
$i_\mathrm{sky}$ (deg) & 87.4068 & 84.1822 & 57.3388 & 78.5862 & 92.4849 \\
$\Delta\Omega_\mathrm{sky}$ (deg)\tablenotemark{a} & 6.02021 & -30.5678 & -9.1363 & 5.06333 & 15.0869 \\
$M$ (deg) & -338.004 & -71.0875 & 72.6743 & -224.326 & -109.464 \\
$M_p$ ($M_{\rm Jup}$) & 0.499494 & 0.231313 & 0.333933 & 0.544041 & 0.164811  \\
\multicolumn{6}{l}{\textit{System}} \\
$i_{\rm mut}$ (deg) & 6.89342 & 30.9392 & 33.3387 & 12.1123 & 15.174 \\
$I$ (deg)\tablenotemark{b} & 89.2356 & 88.0464 & 81.585 & 85.4917 & 91.3988 \\
$\Delta \ln \mathcal{P}$ & 0 & -10.5 & -16.3 & -5.8 & -17.9 \\
$P_\Omega$ (yr)\tablenotemark{d} & 30.4 & 71.7 & 48.3 & 47.2 & 50.0 \\
\enddata
\tablecomments{All orbital elements are osculating elements at the beginning of the $N$-body integration, chosen to be 0.1 day before the start of the first observed transit of TOI-2535\,b.}
\tablenotetext{a}{Relative longitude of nodes, measured in the sky plane.}
\tablenotetext{b}{Inclination of the invariable plane relative to the sky plane.}
\tablenotetext{c}{Nodal precession period, derived from fitting the predictions for $\Omega_\mathrm{inv}$ from the $N$-body model.}
\end{deluxetable}

\end{document}